\documentclass[conference]{IEEEtran}
\IEEEoverridecommandlockouts
\usepackage{cite}
\usepackage{amsmath,amssymb,amsfonts}
\usepackage{algorithmic}
\usepackage{graphicx}
\usepackage{textcomp}
\usepackage{xcolor}

\usepackage[ruled,vlined]{algorithm2e}
\usepackage{makecell}
\usepackage{multirow}
\usepackage{stackengine}
\usepackage{bm}

\usepackage{enumitem}
\usepackage{subfigure}
\usepackage{caption}
\usepackage{url}

\let\oldnl\nl
\newcommand{\nonl}{\renewcommand{\nl}{\let\nl\oldnl}}

\newtheorem{myExample}{Example}
\newtheorem{myDef}{Definition}
\newtheorem{myTheorem}{Theorem}
\newtheorem{myLemma}{Lemma}
\newcommand{\tabincell}[2]{\begin{tabular}{@{}#1@{}}#2\end{tabular}}
\SetKw{Continue}{continue}
\SetKw{Break}{break}
\SetKwRepeat{Do}{do}{while}

\usepackage{setspace}
\makeatletter
\newcommand\figcaption{\def\@captype{figure}\caption}
\newcommand\tabcaption{\def\@captype{table}\caption}
\makeatother

\renewcommand{\IEEEQED}{\IEEEQEDopen}
\def\IEEEproofindentspace{2\parindent}
\renewcommand{\IEEEproofindentspace}{10pt}

\def\BibTeX{{\rm B\kern-.05em{\sc i\kern-.025em b}\kern-.08em
    T\kern-.1667em\lower.7ex\hbox{E}\kern-.125emX}}
\begin{document}

\title{Scalable Community Search with Accuracy Guarantee on Attributed Graphs}

\author{
		\IEEEauthorblockN{Yuxiang Wang\textsuperscript{1}, Shuzhan Ye\textsuperscript{1}, Xiaoliang Xu\textsuperscript{1}, Yuxia Geng\textsuperscript{1}, Zhenghe Zhao\textsuperscript{1}, Xiangyu Ke\textsuperscript{2}, Tianxing Wu\textsuperscript{3}}
		\IEEEauthorblockA{\textsuperscript{1} \textit{Hangzhou Dianzi University, China}, 
			\textsuperscript{2} \textit{Zhejiang University, Hangzhou, China},
			\textsuperscript{3} \textit{Southeast University, Nanjing, China}}
		\{lsswyx,yeshuzhan123,xxl,yuxia.geng,zhaozh\}@hdu.edu.cn, xiangyu.ke@zju.edu.cn, tianxingwu@seu.edu.cn
	}
\maketitle

\begin{abstract}
Given an attributed graph $G$ and a query node $q$, \underline{C}ommunity \underline{S}earch over \underline{A}ttributed \underline{G}raphs (CS-AG) aims to find a structure- and attribute-cohesive subgraph from $G$ that contains $q$. Although CS-AG has been widely studied, they still face three challenges. (1) Exact methods based on graph traversal are time-consuming, especially for large graphs. Some tailored indices can improve efficiency, but introduce nonnegligible storage and maintenance overhead. (2) Approximate methods with a loose approximation ratio only provide a coarse-grained evaluation of a community's quality, rather than a reliable evaluation with an accuracy guarantee in runtime. (3) Attribute cohesiveness metrics often ignores the important correlation with the query node $q$. We formally define our CS-AG problem atop a $q$-centric attribute cohesiveness metric considering both textual and numerical attributes, for $k$-core model on homogeneous graphs. We show the problem is NP-hard. To solve it, we first propose an exact baseline with three pruning strategies. Then, we propose an index-free sampling-estimation-based method to quickly return an approximate community with an accuracy guarantee, in the form of a confidence interval. Once a good result satisfying a user-desired error bound is reached, we terminate it early. We extend it to heterogeneous graphs, $k$-truss model, and size-bounded CS. Comprehensive experimental studies on ten real-world datasets show its superiority, e.g., at least 1.54$\times$ (41.1$\times$ on average) faster in response time and a reliable relative error (within a user-specific error bound) of attribute cohesiveness is achieved.
\end{abstract}

\section{Introduction}
\label{sec:intro}
Recently, it is prevalent to observe that many large-scale and real-world attributed graphs have emerged in various domains, e.g., social networks, collaboration networks, and knowledge graphs \cite{Zhang2019,Fang2017,Huang2017,Sun2020,Liu2020}. In such graphs, nodes represent entities with attributes and edges represent the relationship between entities. Given an attributed graph $G$ and a query node $q$, community search (CS) aims to find a cohesive community $H\subseteq G$ that contains $q$. It is widely recognized that CS is important for many real-life applications \cite{Miao2022,Fang2020}, such as event planning \cite{Sozio2010}, biological data analysis \cite{Dudley2011,PesantezCabrera2019}, and recommendation \cite{Xu2022,Wang2022a,Fang2020}. For example, one can input a specific disease-related gene to find a community of similar genes from a biological network (e.g., GEO \cite{Clough2016}) that helps revealing the hidden causes of diseases; a system can recommend movies for a user by taking one of her favorite movies as a query and return a cohesive community of similar movies from IMDB.



\begin{figure*}
\setlength{\abovecaptionskip}{0.1cm}
\vspace{-0.4cm}
  \centering
  \includegraphics[scale=0.46]{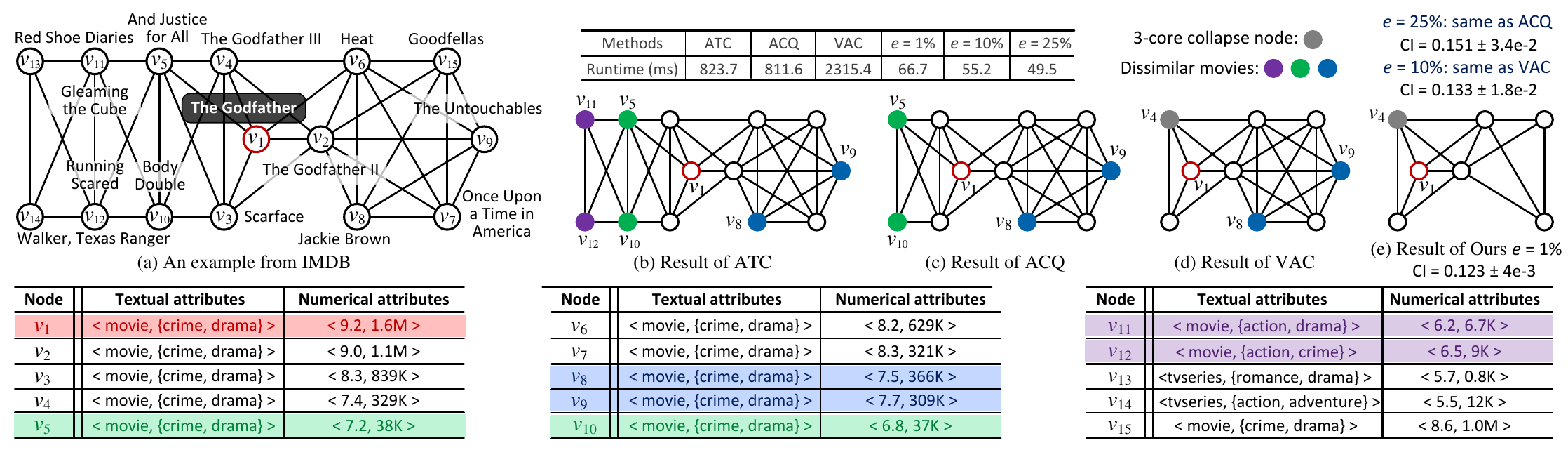}
  \caption{An example of CS: (a) A snapshot of IMDB with attributes at the bottom. (b)-(e) Different results of four methods.}
  \label{fig:Fig1}
\vspace{-0.7cm}
\end{figure*}

Cohesiveness is an important metric for measuring a community's quality, which is two-fold, including structure and attribute cohesiveness. Many models like $k$-core \cite{Sozio2010,Barbieri2015,Cui2014}, $k$-truss \cite{Huang2015,Huang2014,Akbas2017,Liu2020}, and $k$-clique \cite{Cui2013,Tsourakakis2013} are proposed to measure the structure cohesiveness. In terms of the ability to measure structure cohesiveness, \cite{fang2020survey} ranks them as $k$-core $\preceq$ $k$-truss $\preceq$ $k$-clique. Here, $A\preceq B$ indicates the model $B$ is more cohesive than model $A$. As the structure cohesiveness increases, the algorithmic computational efficiency decreases as $k$-core $\succeq$ $k$-truss $\succeq$ $k$-clique \cite{fang2020survey}. Users may choose an appropriate model according to their actual demands. Attribute cohesiveness is another metric for enhancing a community's quality. Given an attribute cohesiveness metric, one can form an optimization problem to find the most attribute-cohesive community. Essentially, existing CS on attributed graphs requires users first to define an appropriate community model and an attribute cohesiveness metric, then design an exact or approximate CS algorithm to return a community of interest \cite{Fang2016,Zhu2020,Liu2020,Huang2017}. Figure \ref{fig:Fig1}(a) shows an example on IMDB. Each node represents an audiovisual work with attributes provided at the bottom (i.e., $\langle$type, $\{$genres$\}\rangle$, $\langle$average rating, \# ratings$\rangle$). Each edge shows that two works have a common actor. Suppose a user likes \textit{The Godfather} (i.e., $v_1$), it is easy to explore a high-rating community of crime drama movies for her by running a CS algorithm given the query as $v_1$.


\noindent\textbf{Challenges and solutions.} We next use the example of movie community search mentioned above to illustrate the challenges faced by existing methods and our solutions.



\noindent\textit{\underline{Challenge I}: How to design a metric to better measure a community's attribute cohesiveness?} Figure \ref{fig:Fig1}(b)-(d) shows the results of three representative CS methods on IMDB given $q=v_1$ (\textit{The Godfather)} and $k$-core ($k=3$) as the community model (a $k$-core is a subgraph of which each node has a degree $\geq k$). ATC \cite{Huang2017} uses the weighted sum of the contribution of $q$'s attributes as the metric. An attribute $a$'s contribution is defined as $\frac{|V_a\cap V_H|^2}{|V_H|}$, where $V_a$ is a set of nodes that have attribute $a$ and $V_H$ is the node set of a community $H$. It excludes $v_{13}$ and $v_{14}$ with different type (TV series) and genres and returns an $H$ with the largest weighted sum of $\frac{13^2}{13}+\frac{12^2}{13}+\frac{12^2}{13}=35.2$ in Figure \ref{fig:Fig1}(b), where the contributions of $\langle$movie,$\{$crime,drama$\}\rangle$ are $\frac{13^2}{13}$, $\frac{12^2}{13}$, and $\frac{12^2}{13}$. Since this metric is tailored for textual attributes based on equality matching, it involves six movies with dissimilar numerical attributes to $q$ (colored nodes), e.g., two low-rating action movies $v_{11}$ and $v_{12}$. Figure \ref{fig:Fig1}(c) shows the result of ACQ \cite{Fang2016} using the \# shared attributes as the metric. It increases the \# shared attributes from 1 (movie) to 3 ($\langle$movie,$\{$crime,drama$\}$) by deleting $v_{11}$ and $v_{12}$. It also includes four dissimilar movies to $q$, as the \# shared attributes is only valid for textual attributes. VAC \cite{Liu2020} is the state-of-the-art work that aims to minimize the maximum attribute distance of any two nodes in a community. Since the maximum attribute distance is dominated by the most dissimilar pair of nodes, it is considered a method that optimizes only the worst case in a community but overlooks the similarity of other nodes to $q$. In Figure \ref{fig:Fig1}(d), $v_4$ is the most dissimilar node to $v_1$, however, deleting it would collapse the $k$-core of $v_1$. So, VAC halts at this step, as the worst case cannot be further improved, making an involvement of $v_8$, $v_9$ that are numerically dissimilar to $q$. In a nutshell, \textit{none of them well-support CS on attributed graphs, either because they cannot simultaneously handle textual and numerical attributes, or because they rely on metrics designed from a global perspective (e.g., optimizing the worst case), neglecting the crucial cohesiveness w.r.t. the query node $q$.}

Therefore, we define a $q$-centric attribute distance of a community $H$ as $\delta(H)$ considering both textual and numerical attributes (\textbf{\S \ref{sec:pre_pro}}), based on which we define our CS problem on attributed graphs (CS-AG) and prove it is NP-hard (\textbf{\S \ref{sec:NP}}). We present an exact algorithm with three pruning strategies (\textbf{\S \ref{sec:exact}}) serving as an exact baseline used in our experimental study. 


\vspace{0.1cm}
\noindent\textit{\underline{Challenge II}: How to design an efficient index-free approximate CS algorithm with a reliable accuracy guarantee?} The conventional solution for CS is to design algorithms based on graph traversal, which is time-consuming for large graphs \cite{Fang2020,Miao2022,Liu2020,Huang2017,Fang2016}. To improve efficiency, researchers resort to community-aware indices or approximate algorithms. For index-based solutions, a trade-off between efficiency and space overhead is achieved. However, an index that is comparable to the graph size usually is required \cite{Fang2016}. In dynamic scenarios, the index update or reconstruction would introduce nonnegligible maintenance overhead. For approximate solutions, a trade-off between efficiency and effectiveness is achieved, e.g., \cite{Huang2015,Liu2020} provide $2$-approximation to the exact results via triangle inequality \cite{Kosub2019}. However, \textit{the approximate ratio only provides an upper (often loose) bound of attribute cohesiveness but fails to offer a reliable evaluation of how good a community is in runtime, i.e., what is the relative error of a community's attribute distance w.r.t. that of the ground-truth community?}

Comparing to a tardy exact or an approximate result with a loose approximation ratio, it's more desirable if a method can quickly return an approximate result with a reliable accuracy guarantee \cite{Laptev2012,Chaudhuri2017,Wang2022,Wang2014}. Thus, we present an index-free sampling-estimation-based approximate algorithm to first collect a set of high-quality samples (nodes) based on \textit{Hoeffding Inequality} to form a candidate community. We then estimate its attribute distance as $\delta^*$ with a $1-\alpha$ level confidence interval CI = $\delta^{\star}\pm \varepsilon$ using \textit{Bag of Little Bootstraps}, where $\varepsilon$ is the half width of a CI. Given a user-desired error bound $e$, we guarantee that the relative error of $\delta^{\star}$ w.r.t. the exact $\delta$ is bounded by $e$ (i.e., $|\delta^{\star}-\delta|/\delta\leq e$) if $\varepsilon$ is small enough to satisfy Theorem \ref{th:relative_error}. We early terminate the query once such a good $\delta^{\star}$ is reached. Otherwise, we enlarge the sample and repeat above until Theorem \ref{th:relative_error} holds. Figure \ref{fig:Fig1}(e) shows the result of $e$ = $1\%$ and $1-\alpha$ = $95\%$, it costs only 66 ms (at least 12X faster than others, see the table at the top) to return the same community as that of our exact algorithm (\S \ref{sec:exact}), indicating a relative error of 0\% ($\delta^{\star}$ = $\delta$ = 0.123). It involves seven similar crime drama movies to $v_1$ with higher rating and more ratings. By varying $e$, we found that the results of $e$ = $10\%$ and $25\%$ are the same as (d) and (c) with bounded relative errors. For example, given the CIs provided above Figure \ref{fig:Fig1}(e), we have a relative error for $e$ = $10\%$ as $\frac{0.133-0.123}{0.123}=8.1\%$. A larger $e$ and smaller $1-\alpha$ usually result in less time for estimation, e.g., 66 ms, 55 ms, and 49 ms for $e=1\%$, $10\%$, and $25\%$. Users may configure them based on their preferences for accuracy and efficiency.

\vspace{0.1cm}
\noindent\textit{\underline{Challenge III}: How to enable a CS algorithm to be scalable for different scenarios?} A CS algorithm usually is tailored for a specific graph type (homogeneous or heterogeneous) given a specific community model ($k$-core, $k$-truss, etc.) with a size-constraint (size-bounded CS \cite{Yao2021}) or not, showing a strong dependence on a specific scenario. \textit{This implies that if we want to adapt a CS algorithm of one scenario to other scenarios, usually we must redesign it partially or even completely.}

Since our approximate CS algorithm consists of two standard steps: sampling and estimation, it is easy to be extended to more general scenarios with lightweight modifications, including heterogeneous graphs (\textbf{\S \ref{sec:hetero}}), size-bounded CS (\textbf{\S \ref{sec:size}}), and other community models (\textbf{\S \ref{sec:models}}). For example, we can simply add a constraint on the sample size in sampling step to support size-bounded CS with an accuracy guarantee.



\vspace{0.1cm}
\noindent\textbf{Contributions.} The main contributions are as follows.
\begin{itemize}[leftmargin=*]
\item We define a CS-AG problem and its approximate version Approx-CS-AG based on a $q$-centric attribute cohesiveness metric in \textbf{\S \ref{sec:pre_pro}}. We prove CS-AG is NP-hard in \textbf{\S \ref{sec:NP}}.
\item We present an exact method to CS-AG with three pruning strategies (\textbf{\S \ref{sec:exact}}) to avoid redundant searches.
\item We propose an index-free sampling-estimation-based approximate CS method to Approx-CS-AG in \textbf{\S \ref{sec:sampling_solution}}, with a reliable runtime accuracy guarantee on attribute cohesiveness. 
\item We extend our approximate CS method to more general scenarios such as CS on heterogeneous graphs, size-bounded CS, and CS with different community models in \textbf{\S \ref{sec:extension}}.
\item We conduct extensive experiments to evaluate our method from: effectiveness and efficiency (\textbf{\S \ref{sec:effective}-\ref{sec:efficiency}}), effect of pruning strategies (\textbf{\S \ref{sec:prune}}), scalability analysis (\textbf{\S \ref{sec:scale}}), case study (\textbf{\S \ref{sec:case}}), and parameter sensitivity (\textbf{\S \ref{sec:parameter}}).
\end{itemize}

\section{Preliminaries and Problems}
\label{sec:pre_pro}

\subsection{Preliminaries}
\label{sec:pre}

\begin{myDef}
\label{def:attributed_graph}
{\rm \textbf{Attributed Graph}}. An attributed graph is defined as $G=(V_G,E_G,A_G)$, where $V_G$ ($E_G$) is the node (edge) set and $A_G$ is the set of attributes associated with nodes. For each node $v\in V_G$, it has a set of attributes $A(v)=\{a_0,\dots,a_m\}$, where $A(v)\subseteq A_G$ and $A(v)_i$ indicates the $i$-th attribute $a_i$ of a node $v$. In this paper, we consider both the textual and numerical attributes of a node, denoted by $A^t(v)\subseteq A(v)$ and $A^{\#}(v)\subseteq A(v)$, respectively.
\end{myDef}

\begin{figure}
\setlength{\abovecaptionskip}{0.1cm}
  \centering
  \includegraphics[scale=0.44]{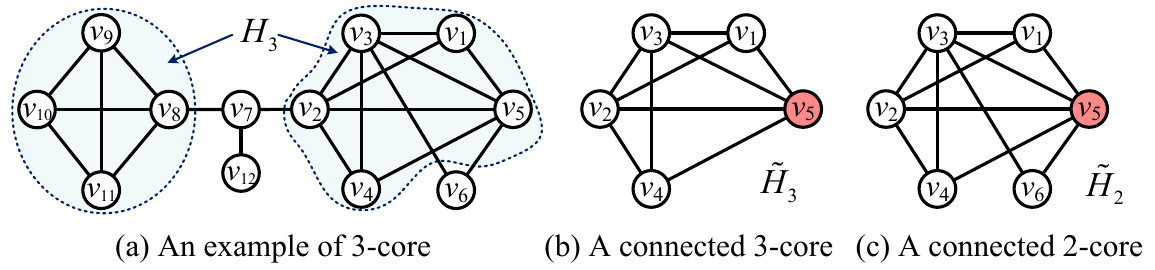}
  \caption{An example of $k$-core and connected $k$-core}
  \label{fig:kcore}
\vspace{-0.6cm}
\end{figure}


\vspace{0.1cm}
\noindent \textbf{Structure cohesiveness.} We first employ $k$-core model \cite{Bonchi2019} to measure the structure cohesiveness of a community, because it is the most efficient model compared to other models \cite{fang2020survey}, i.e., $k$-truss \cite{Huang2015}, $k$-ECC \cite{Hu2016}, $k$-clique \cite{Cui2013}, and it performs well on cohesiveness evaluation \cite{Fang2020,fang2020survey,Li2015,Cui2014,Fang2017}. In \S \ref{sec:extension}, we show how to extend it to other community models.


\vspace{0.1cm}
\begin{myDef}
\label{def:kcore}
{\rm \textbf{$k$-core}}. Given a graph $G$ and a non-negative integer $k$, a $k$-core of $G$ is the largest subgraph $H_k\subseteq G$, such that $\forall v\in H_k$ has a degree at least $k$, i.e., $deg(v,H_k)\geq k$. 
\end{myDef}

\vspace{0.1cm}
In Figure \ref{fig:kcore}(a), $H_1$ is the entire graph as every node has at least one neighbor; $H_2$ is the subgraph excluding $v_{12}$; $H_3$ consists of two components and each node has at least three neighbors. The larger $k$ is, the more cohesiveness $H_k$ has. 



\vspace{0.1cm}
\begin{myDef}
\label{def:ckcore}
{\rm \textbf{Connected $k$-core \cite{Sun2020}}}. Given a graph $G$ and a non-negative integer $k$, a connected $k$-core of $G$ is a connected subgraph $\tilde{H}_k\subseteq H_k\subseteq G$, such that $\forall v\in \tilde{H}_k$, $deg(v,\tilde{H}_k)\geq k$.
\end{myDef}

\vspace{0.1cm}
To incorporate connectivity into a community, similar to \cite{Sun2020,Li2018,Barbieri2015,Fang2016}, we use connected $k$-core as the community model. Given a query node $q$, we aim to find a connected $k$-core containing $q$ as the desired community of $q$. Figure \ref{fig:kcore}(b-c) shows two communities $\tilde{H}_3$ and $\tilde{H}_2$ containing $q=v_5$. 





\vspace{0.1cm}
\noindent \textbf{Attribute cohesiveness.} We consider two types of attributes: \textit{textual} and \textit{numerical}. Given two nodes $u,v\in V_G$, they are similar in textual attributes if their textual attributes have substantial overlap \cite{Liu2020}. We use \textit{Jaccard distance} to measure their textual attribute distance, denoted by $f^t(u,v)=1-\frac{|A^t(u) \cap A^t(v)|}{| A^t(u) \cup A^t(v)|}$. The essential of Jaccard distance is the equality matching, so that the higher the ratio of equally matched textual attributes, the smaller the $f^t(u,v)$. Unlike textual attributes, equality matching is often meaningless for numerical attributes. Many alternative choices are available, such as Manhattan distance and Euclidean distance. In this paper, we compute the numerical attribute distance based on Manhattan distance $f^{\#}(u,v)=\frac{\sum_{i=1}^{m} |Z(A^{\#}(u)_i) - Z(A^{\#}(v)_i)|}{m}$, where we use $Z(\cdot)$ to normalize $A^{\#}(\cdot)$ to [0,1], thereby eliminating the dimensional influence. We next compute the composite attribute distance between two nodes as $f(u,v)=\gamma \cdot f^t(u,v) + (1-\gamma) \cdot f^{\#}(u,v)$, where the parameter $\gamma\in[0,1]$ is configured as a balance factor to adjust user preferences for different types of attribute cohesiveness. Its worth mentioning that we may have other options to combine two types of attribute distances. For example, if two types of attributes are vectorized as a unified high-dimensional vector (i.e., embeddings), we can simply measure the attribute distance by computing cosine similarity of attribute vectors \cite{Wang2024}. Using the composite distance, we can measure how similar a node is to a query node $q$. Intuitively, the more nodes are similar to $q$, the more attribute cohesiveness w.r.t. $q$ it is. We next define the $q$-centric \textit{attribute distance} of a community $\tilde{H}_k$ as follows. 




\vspace{0.1cm}
\begin{myDef}
\label{def:attribute_dis}
\textbf{Attribute distance of $\tilde{H}_k$}. Given a community $\tilde{H}_k$ containing $q$, its $q$-centric attribute distance is defined as the average composite attribute distance to $q$ over all nodes $\in \tilde{V}_k\setminus q$, i.e., $\delta(\tilde{H}_k)=\frac{\sum_{\forall u\in \tilde{V}_k\setminus q} f(u,q)}{|\tilde{V}_k|-1}$, $\tilde{V}_k$ is $\tilde{H}_k$'s node set.
\end{myDef}

\subsection{Problem Definition}
\label{sec:pro}
\noindent\textbf{Problem 1.} (\textbf{CS-AG problem}) Given an attributed graph $G=(V_G,E_G,A_G)$, a query node $q \in V_G$, and a parameter $k>0$, the \underline{C}ommunity \underline{S}earch over \underline{A}ttributed \underline{G}raphs (CS-AG) returns a subgraph $\tilde{H}_k\subseteq G$ satisfying the following properties:
\begin{itemize}
\item \textbf{Query participation.} $\tilde{H}_k$ contains $q$, i.e., $q\in \tilde{V}_k$;
\item \textbf{Structure cohesiveness.} $\tilde{H}_k$ is a connected $k$-core;
\item \textbf{Attribute cohesiveness.} $\tilde{H}_k$ has the smallest $\delta(\tilde{H}_k)$.
\end{itemize}

We prove that CS-AG problem is NP-hard in \S \ref{sec:NP} and provide an exact baseline used in experimental study (\S \ref{sec:exp}). Since it is time-consuming for large graphs, we define an approximate version as Approx-CS-AG, and present a sampling-estimation-based method with an accuracy guarantee in \S \ref{sec:sampling_solution}. 

\vspace{0.1cm}
\noindent\textbf{Problem 2.} (\textbf{Approx-CS-AG problem}) Given an attributed graph $G=(V_G,E_G,A_G)$, a query node $q \in V_G$, a parameter $k>0$, a user-specific error bound $e$, and a confidence level $1-\alpha$, Approx-CS-AG returns an approximate community $\tilde{H}^\star_k\subseteq G$ containing $q$ that satisfies: (1) the attribute distance $\delta(\tilde{H}_k)$ (shorten as $\delta$) of the exact community $\tilde{H}_k$ is covered by a confidence interval of the attribute distance $\delta(\tilde{H}^\star_k)$ (shorten as $\delta^\star$) of $\tilde{H}^\star_k$ (Eq. \ref{eq:CI}), and (2) the relative error of $\delta^\star$ w.r.t. $\delta$ is bounded by the user-specific error bound $e$ (Eq. \ref{eq:error_rate}).
\begin{equation}
\label{eq:CI}
{\rm Pr}[\delta^\star-\varepsilon\leq \delta\leq \delta^\star+\varepsilon]=1-\alpha
\end{equation}
\begin{equation}
\label{eq:error_rate}
|\delta^\star-\delta|/\delta\leq e
\end{equation}

We use a confidence interval CI = $\delta^\star\pm \varepsilon$ to quantify the quality of $\delta^\star$, which states that $\delta$ is covered by a range $\delta^\star\pm \varepsilon$ with a probability of $1-\alpha$. The half width $\varepsilon$ is called the Margin of Error (MoE). The smaller $\varepsilon$ shows a higher quality of $\delta^\star$ \cite{Wang2022}. In \S \ref{sec:acc_guarantee}, we prove that the accuracy guarantee $|\delta^\star-\delta|/\delta\leq e$ is ensured if $\varepsilon\leq \delta^\star\cdot e/(1+e)$ (Theorem \ref{th:relative_error}). 


\section{Hardness Analysis}
\label{sec:NP}
We first define CS-AG's decision version as $\rho$CS-AG. Next, we outline the idea of reduction from a NP-hard R-set Maximum-Weight Connected Subgraph (R-MWCS) problem \cite{ElKebir2014} to $\rho$CS-AG, and show the proof in Theorem \ref{th:nphard}.



\vspace{0.1cm}
\noindent\textbf{Problem 3.} (\textbf{$\rho$CS-AG problem}) Given an attributed graph $G=(V_G, E_G, A_G)$, a query node $q\in V_G$, and two parameters $k$, $\rho\geq 0$, the problem checks if there exists a connected $k$-core $\tilde{H}_k\subseteq G$ containing $q$ with the attribute distance $\delta(\tilde{H}_k)\leq \rho$. 

Generally, the decision version of a problem is easier than (or the same as) it's optimization version \cite{Kleinberg2006}. Thus, if $\rho$CS-AG is NP-hard, then CS-AG is NP-hard. To achieve this, we first prove the decision version of R-MWCS (called $\tau$R-MWCS, defined below) is NP-hard. Then, we reduce $\tau$R-MWCS to our $\rho$CS-AG to complete the proof.  

\vspace{0.1cm}
\noindent\textbf{Problem 4.} (\textbf{$\tau$R-MWCS problem}). Given a graph $G=(V_G, E_G)$, a node set $R\subseteq V_G$, and node weights $w: V_G\rightarrow \mathbb{R}$ ($w(v)$ indicates the weight of $v\in V_G$, which could be any real number). It finds a connect subgraph $H$ that contains $R$, satisfying the graph weight $w(H)=\sum_{v\in V_H} w(v)\leq \tau$.

\vspace{0.1cm}
\begin{myTheorem}
\label{th:rmwcs_nphard} 
The $\tau${\rm R-MWCS} problem is NP-hard.
\end{myTheorem}

\vspace{0.1cm}
\begin{IEEEproof}
We reduce the NP-hard Steiner tree problem (decision version, shorten as STP) \cite{STP,Byrka2010} to $\tau${\rm R-MWCS}. Given a graph $G=(V_G,E_G)$, a node set $S\subseteq V_G$, and a number $\tau\in \mathbb{N}$, STP checks if there is a tree that contains $S$ and includes at most $\tau$ edges. We complete the proof following the similar logic of \cite{ElKebir2014}. First, we construct an instance graph $G'$ via a polynomial time transformation from $G$. For each edge $(v,u)\in E_G$, we introduce a middle node $w$ to split $(v,u)$ into two edges $(v,w)$, $(w,u)\in E_{G'}$, and assign weights on $v,u,w$ as $0,0,1$. We next prove that the instance of STP is a Yes-instance iff the instance of $\tau${\rm R-MWCS} is a Yes-instance.

\vspace{0.1cm}
\noindent$(\Rightarrow)$ If a tree $T\subseteq G$ is a solution of STP, then it has $|E_T|\leq \tau$. Since we split each edge in $E_T$ into two edges with a new middle node having a weight of 1 and other two original nodes having a weight of 0, the corresponding tree $T'\subset G'$ must have $w(T')\leq \tau$. $T'$ is also a connected subgraph of $G'$, so that $T'$ is a Yes-instance of $\tau${\rm R-MWCS}.

\vspace{0.1cm}
\noindent$(\Leftarrow)$ If a subgraph $H'\subseteq G'$ is a solution of $\tau${\rm R-MWCS}, then it has $w(H')\leq \tau$. Since $H'$ is connected, a spanning tree $T'\subseteq H'$ still satisfies $w(T')\leq \tau$. For each node with weight of 1 in $T'$, it corresponds to an edge $(v,u)\in E_G$. This implies that the edges in the corresponding $T\subseteq G$ are at most $\tau$. Thus, $T$ is a Yes-instance of STP.
\end{IEEEproof}


\vspace{0.1cm}
\begin{myTheorem}
\label{th:nphard}
The $\rho${\rm CS-AG} problem is NP-hard.
\end{myTheorem}

\vspace{0.1cm}
\begin{IEEEproof}
We reduce $\tau$R-MWCS problem to $\rho$CS-AG. Given a graph $G=(V_G, E_G,A_G)$, we first construct an instance graph $G^*$ via a polynomial time transformation from $G$. First, we add all nodes and edges of $G$ into $G^*$. Second, we add additional $k|V_G|$ nodes into $G^*$ and assign a unique node ID for all the $(k+1)|V_G|$ nodes: (1) nodes from $V_G$ have IDs $\in [0,|V_G|-1]$ and (2) additional $k|V_G|$ nodes have IDs $\in [|V_G|,(k+1)|V_G|-1]$. Third, we add additional edges for pair-wise nodes $u$ and $v$, if $u.$ID $\%|V_G|=v.$ID $\%|V_G|$ (e.g., nodes with IDs of $\{1,|V_G|+1,\dots,k|V_G|+1\}$ have edges). Hence, we ensure that each node in $G^*$ has at least $k$ neighbors. Finally, given a node set $R$ with only one query node $q$, we assign weights on both $G$ and $G^*$ as follows: (1) $\forall v\in G$ has $w(v)=f(v,q)-\rho$, where $f(v,q)$ is the composite attribute distance between $v$ and $q$. (2) $\forall v\in G^*$ with ID $\in [0,|V_G|-1]$ has $w^*(v)=w(v)+\rho=f(v,q)$, which equals to $v$'s original composite attribute distance to $q$. (3) $\forall u\in G^*$ with ID $\in [|V_G|,(k+1)|V_G|-1]$, we assign $u$ with a same weight $w^*(u)=\rho$ as its $f(u,q)$. In summary, considering $G^*$, those nodes inherited from $G$ have weights as their original $f(v,q)$ and others are configured with the same weight $\rho$. Given $G^*$ with the query node $q$ and $\rho\in[0,1]$, and $G$ with $R=\{q\}$ and $\tau=-\rho$, we show that the instance of $\tau$R-MWCS is a Yes-instance iff the instance of $\rho$CS-AG is a Yes-instance.


\vspace{0.1cm}
\noindent$(\Rightarrow)$ If $H\subseteq G$ is a solution to the $\tau$R-MWCS problem that satisfies $w(H)=\sum_{v\in V_H} w(v)\leq -\rho$, then the induced graph formed by nodes of $V_H\cup V'$ is a connected $k$-core $\tilde{H}_k$, where $V'$ includes $k|V_H|$ nodes with IDs $\in [|V_G|,(k+1)|V_G|-1]$ that have edges with nodes in $V_H$. Since $w(H)=\sum_{v\in V_H} w(v)\leq -\rho$, we have $\sum_{v\in V_H\setminus q} w(v)\leq 0$ (as $w(q)=-\rho$) and $\sum_{v\in V_H\setminus q} (w(v)+\rho)=\sum_{v\in V_H\setminus q} w^*(v)\leq \rho$. Thus, $\delta(\tilde{H}_k)=\frac{\sum_{v\in V_H\setminus q} w^*(v) + \sum_{u\in V'}\rho}{|V_H\cup V'\setminus q|}\leq \rho$. So, $\tilde{H}_k$ is a Yes-instance of $\rho$CS-AG if $H$ is a Yes-instance of the $\tau$R-MWCS.

\vspace{0.1cm}
\noindent$(\Leftarrow)$ Assume a connected $k$-core $\tilde{H}_k$ containing $q$ is a solution to the $\rho$CS-AG problem. $\tilde{H}_k$ involves part of nodes with IDs $\in [0,|V_G|-1]$ (i.e., $V_H\subseteq V_{\tilde{H}_k}$) and the rest of nodes with IDs $\in [|V_G|,(k+1)|V_G|-1]$ (denoted by $V'=V_{\tilde{H}_k}$ $\setminus$ $V_H$). Given $\delta(\tilde{H}_k)=\frac{\sum_{v\in V_H\setminus q} w^*(v) + \sum_{u\in V'}\rho}{|V_H\cup V'\setminus q|}\leq \rho$, we have $\delta(\tilde{H}_k)-\rho=\frac{\sum_{v\in V_H\setminus q} (w^*(v)-\rho)}{|V_H\cup V'\setminus q|}\leq 0$, i.e.,$\sum_{v\in V_H\setminus q} (w^*(v)-\rho)$ $=$ $\sum_{v\in V_H\setminus q} w(v)$ $\leq$ $0$. Thus, we have $w(H)=\sum_{v\in V_H\setminus q} w(v)+w(q)\leq -\rho=\tau$ (as $w(q)=-\rho$). So, $H$ is a Yes-instance of $\tau$R-MWCS if $\tilde{H}_k$ is a Yes-instance of $\rho$CS-AG.
\end{IEEEproof}


\section{Exact Baseline}
\label{sec:exact}
Given an attributed graph $G$, we present an \texttt{Exact} method to solve CS-AG. Since the ground-truth community must be included in the maximal connected $k$-core $\tilde{H}_k\subseteq G$, we first find the maximal $\tilde{H}_k$ containing $q$ (\S \ref{sec:max_core}). Then, we enumerate all the $\tilde{H}^i_k\subseteq \tilde{H}_k$ containing $q$ with three pruning strategies (\S \ref{sec:enumeration}), and return the one with the smallest $\delta(\cdot)$. 


\subsection{Find the Maximal Connected $k$-core}
\label{sec:max_core}
We have two straightforward ways to obtain the maximal $\tilde{H}_k$: one is the classic \textit{core-decomposition} \cite{batagelj2003m} and another is the \textit{search with expansion} \cite{Fang2020}. For the latter one, we start the search from $q$ and maintain up to $k$ neighbors for each explored node $v$; if $v$ does not have $k$ neighbors, then we delete $v$ and maintain all previously explored nodes' degree up to $k$ if their degree is reduced to be $<k$ after removing $v$. We repeat this until all nodes have been explored. No matter which method is adopted, the essence is the same that is to recursively remove nodes with degree $<k$ from the connected component of $q$. We implement the former one in \texttt{Exact}.

\subsection{Enumeration with Pruning Strategies}
\label{sec:enumeration}
\begin{figure}
\setlength{\abovecaptionskip}{0.2cm}
  \centering
  \includegraphics[scale=0.41]{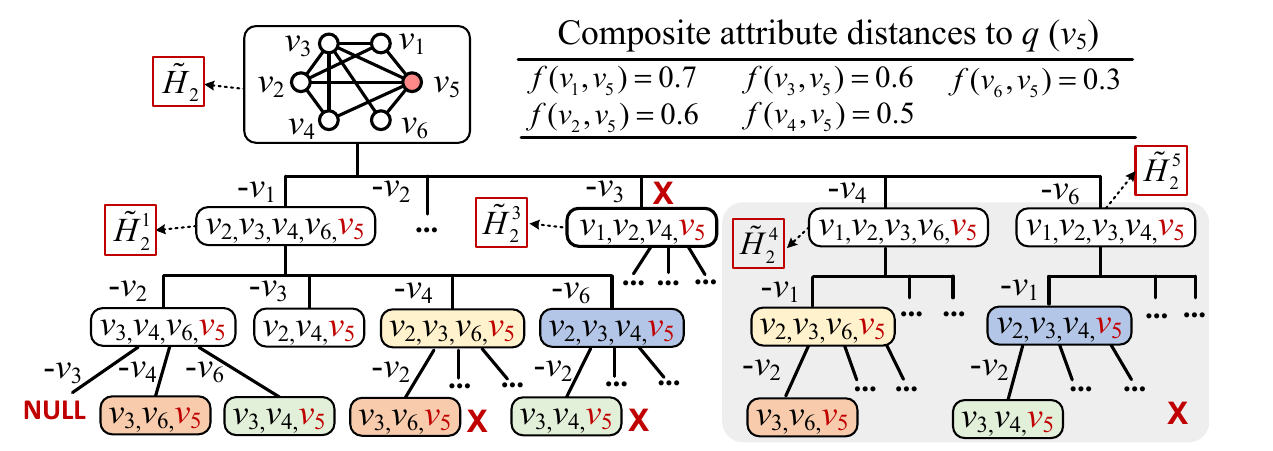}
  \caption{A search tree for the $\tilde{H}_2$ in Figure \ref{fig:kcore} (c)}
  \label{fig:prune}
\vspace{-0.6cm}
\end{figure}

Given the maximal $\tilde{H}_k\subseteq G$, we continuously peel nodes from it to form a new candidate community $\tilde{H}^i_k\subseteq \tilde{H}_k$ and check its attribute distance. This enumeration is represented as a search tree, where each state indicates a $\tilde{H}^i_k$ and the root $\tilde{H}^0_k$ is initialized as $\tilde{H}_k$. If $\tilde{H}_k$ includes $n$ nodes, then we can delete nodes iteratively (except $q$) to generate $n-1$ substates. Figure \ref{fig:prune} illustrates a search tree starting from the $\tilde{H}_2$ in Figure \ref{fig:kcore} (c), given $q=v_5$. For example, $\tilde{H}^1_2\subseteq \tilde{H}_2$ is obtained by deleting $v_1$. A full enumeration is computationally expensive for a large $\tilde{H}_k$, but not all $\tilde{H}^i_k\subseteq \tilde{H}_k$ should be visited and some of them can be pruned safely to improve the efficiency.


\vspace{0.1cm}
\noindent \textbf{Prune for duplicate states.} In Figure \ref{fig:prune}, we show duplicate states with the same color, e.g., the green states are generated from $\tilde{H}_2$ by deleting $\{v_1,v_2,v_6\}$. A simple way to avoid duplicate states is to record all the visited states and check if each new state has been visited before. The main drawback is that it requires extra memory overhead to maintain all visited states. To handle this, we present our first pruning strategy based on \textit{priority enumeration}. More precisely, we enumerate states in a  partial order w.r.t. every node's composite attribute distance to $q$ (i.e., $f(\cdot,q)$), so that we quickly can decide to prune a state by simply evaluating $f(\cdot,q)$ between the node to be deleted and $q$, without maintaining visited states.

\vspace{0.1cm}
\noindent \underline{(1) Priority enumeration.} Given an arbitrary state, we enumerate its substates in a DFS manner by deleting nodes in descending order of their composite attribute distance $f(\cdot,q)$ to $q$. In Figure \ref{fig:prune}, we first enumerate the state $\tilde{H}^1_2$ by deleting $v_1$ as $f(v_1,q)=0.7$ is larger than other nodes' distance to $q$ (distance information are provided on the top of Figure \ref{fig:prune}).

\vspace{0.1cm}
\begin{myLemma}
\label{lemma:1}
Given a state $\tilde{H}^i_k$ includes two nodes $v_x$ and $v_y$ with $f(v_x,q)>f(v_y,q)$, it must exists a substate $\tilde{H}^j_k\subseteq \tilde{H}^i_k$ generated by successively deleting $v_x$ and $v_y$ from $\tilde{H}^i_k$.
\end{myLemma}

\vspace{0.1cm}
This lemma holds as we enumerate states by deleting nodes in descending order of $f(\cdot,q)$ (i.e., priority enumeration). For example, $\tilde{H}^1_2$ (in Figure \ref{fig:prune}) has a substate with nodes $\{v_3,v_4,v_5\}$ after deleting $v_2$, $v_6$ in turn ($f(v_2,q)>f(v_6,q)$).

\vspace{0.1cm}
\noindent \underline{(2) Prune based on $f(\cdot,q)$.} Given a state $\tilde{H}^i_k$ that is generated from its parent state $\tilde{H}^p_k$ by deleting a node $v_y\in \tilde{H}^p_k$. Suppose we try to enumerate a substate $\tilde{H}^j_k\subseteq \tilde{H}^i_k$ by deleting a current node $v_x\in \tilde{H}^i_k$, then we can prune $\tilde{H}^j_k$ in the following cases.

\vspace{0.1cm}
\noindent $\bullet$ \textit{Case 1}. Let us consider a simple case, where deleting $v_x$ does not result in other nodes from $\tilde{H}^i_k$ having a degree less than $k$, i.e., no other nodes would be removed after deleting $v_x$. In this case, we can prune $\tilde{H}^j_k$ according to Theorem \ref{th:1}.


\vspace{0.1cm}
\begin{myTheorem}
\label{th:1} 
Given a state $\tilde{H}^i_k\subseteq \tilde{H}^p_k$ that is obtained by deleting a node $v_y\in \tilde{H}^p_k$ and a substate $\tilde{H}^j_k\subseteq \tilde{H}^i_k$ obtained by deleting only one node $v_x\in \tilde{H}^i_k$. If $f(v_x,q)>f(v_y,q)$, then we can prune $\tilde{H}^j_k$ and its subsequent sub-states.
\end{myTheorem}

\vspace{0.1cm}
\begin{IEEEproof}
\label{proof:th1}
According to Lemma \ref{lemma:1}, a previously visited state must exist that is caused by successively deleting $v_x$ and $v_y$ from $\tilde{H}^p_k$ because $f(v_x,q)>f(v_y,q)$. This state is duplicated with $\tilde{H}^j_k\subseteq \tilde{H}^p_k$ generated by successively deleting $v_y$ and $v_x$. Thus, we can directly prune $\tilde{H}^j_k$ if $f(v_x,q)>f(v_y,q)$.
\end{IEEEproof}


\vspace{0.1cm}
\begin{myExample}
\label{example:1}
Given the yellow state (in Figure \ref{fig:prune}) including nodes $\{v_2,v_3,v_6,v_5\}$ that is obtained from $\tilde{H}^1_2$ by deleting $v_4$, $v_2$ is the next node to be deleted to get a substate. Since $f(v_2,v_5)>f(v_4,v_5)$, it exists a duplicate state that has been visited before, i.e., the one obtained by successively deleting $v_2$, $v_4$ from $\tilde{H}^1_2$. So, we can prune this substate.
\end{myExample}

\vspace{0.1cm}
\noindent $\bullet$ \textit{Case 2}. Let us consider a more general case where deleting $v_x$ results in other nodes from $\tilde{H}^i_k$ having a degree less than $k$, i.e., additional nodes would be removed when $v_x$ is deleted. In this case, we can prune $\tilde{H}^j_k$ according to Theorem \ref{th:2}.

\vspace{0.1cm}
\begin{myTheorem}
\label{th:2} 
Given a state $\tilde{H}^i_k\subseteq \tilde{H}^p_k$ that is obtained by deleting a node $v_y\in \tilde{H}^p_k$ and a substate $\tilde{H}^j_k\subseteq \tilde{H}^i_k$ obtained by deleting multiple nodes besides $v_x$, i.e., $\tilde{H}^j_k=\tilde{H}^p_k\setminus \{v_y,v_x,\cdots,v_m\}$, where $v_m$ is node with the maximal $f(\cdot,q)$ among all deleted nodes. If $f(v_m,q)>f(v_y,q)$, then we can prune $\tilde{H}^j_k$ and its subsequent substates.
\end{myTheorem}

\vspace{0.1cm}
\begin{IEEEproof}
\label{proof:th2}
Similar to Theorem \ref{th:1}, a previously visited state must exist that is caused by successively deleting $v_m$ and other nodes from $\tilde{H}^p_k$, as $f(v_m,q)>f(v_y,q)$. It is duplicated with the $\tilde{H}^j_k$ generated by successively deleting $\{v_y,v_x,\cdots,v_m\}$ from $\tilde{H}^p_k$. So, we can prune it when $f(v_m,q)>f(v_y,q)$.
\end{IEEEproof}

Since Case 1 is a special case of Case 2 when $v_x=v_m$, we only apply Theorem \ref{th:2} in our implementation of \texttt{Exact}.


\vspace{0.1cm}
\noindent \textbf{Prune for unnecessary states.} We say a state is unnecessary to visit if the optimal community definitely can be visit before this state, so that we can directly prune this state.

\vspace{0.1cm}
\begin{myTheorem}
\label{th:3}
Given a state $\tilde{H}^i_k$ with an attribute distance $\delta(\tilde{H}^i_k)$, the substates that are generated by deleting nodes with $f(\cdot,q)\leq \delta(\tilde{H}^i_k)$ are unnecessary to visit and can be pruned.
\end{myTheorem}

\vspace{0.1cm}
\begin{IEEEproof}
\label{proof:th3}
Suppose we enumerate a new substate of $\tilde{H}^i_k$ by deleting a node $v_y\in \tilde{H}^i_k$ with $f(v_y,q)\leq \delta(\tilde{H}^i_k)$. If $v_y$'s deletion only causes nodes $v_x$ with $f(v_x,q)\leq \delta(\tilde{H}^i_k)$ to be deleted, then this new substate's attribute distance must $\geq \delta(\tilde{H}^i_k)$, indicating it is not the optimal community. This implies that if the optimal community exists, then at least one node $v_x$ with $f(v_x,q)>\delta(\tilde{H}^i_k)$ would be deleted recursively after $v_y$ is deleted. According to Lemma \ref{lemma:1}, a state $\tilde{H}^j_k\subseteq \tilde{H}^i_k$ generated by successively deleting $v_x$ and $v_y$ must be visited previously as $f(v_x,q)$ $>$ $\delta(\tilde{H}^i_k)$ $\geq$ $f(v_y,q)$. So, we can prune $\tilde{H}^j_k$ based on Theorem \ref{th:1}-\ref{th:2}. In summary, substates generated by deleting nodes with $f(\cdot,q)\leq \delta(\tilde{H}^i_k)$ can be pruned.
\end{IEEEproof}

\vspace{0.1cm}
According to Theorem \ref{th:3}, we only need to enumerate substates of $\tilde{H}^i_k$ by deleting those nodes with $f(\cdot,q)> \delta(\tilde{H}^i_k)$.

\vspace{0.1cm}
\begin{myExample}
\label{example:2}
In Figure \ref{fig:prune}, states $\tilde{H}^4_2$ and $\tilde{H}^5_2$ are not duplicated with any previously visited states, but they are unnecessary to visit. Note that, the attribute distance of $\tilde{H}_2$ as $\delta(\tilde{H}_2)=\frac{0.7+0.6+0.6+0.5+0.3}{5}=0.54$. According to Theorem \ref{th:3}, the optimal community must exclude one node $v_x\in \{v_1,v_2,v_3\}$ ($f(v_x,v_5)>0.54$) and it must be visited before $\tilde{H}^4_2$ and $\tilde{H}^5_2$, as $f(v_x,v_5)>f(v_4,v_5),f(v_6,v_5)$ (Lemma \ref{lemma:1}).
\end{myExample}

\vspace{0.1cm}
\noindent \textbf{Prune for unpromising states.} Another case where we can prune is when the lower bound of $\delta(\cdot)$ for all substates of $\tilde{H}^i_k$ is larger than the optimal $\delta^*(\cdot)$ so far. This implies that we cannot find a better community by digging deeper from $\tilde{H}^i_k$.




\vspace{0.1cm}
\noindent \underline{(1) Compute the lower bound $\underline{\delta}(\cdot)$.} Since a $(k+1)$-clique is the smallest $k$-core, each state (represents a connected $k$-core) in a search tree must have at least $k+1$ nodes ($q$ and other $k$ nodes). So, we obtain $k$ nodes with the smallest $f(\cdot,q)$ from the current state $\tilde{H}^i_k$ (Eq. \ref{eq:min_k1}) and use the average $f(\cdot,q)$ of the $k$ nodes as the lower bound of attribute distance $\underline{\delta}(\cdot)$ (Eq. \ref{eq:lowerbound}).
\begin{equation}
\label{eq:min_k1}
V_{\min}=\{\mathop{\arg\min}_{v\in \tilde{H}^i_k} f(v,q)\}\quad s.t.\ |V_{\min}|=k
\end{equation}
\begin{equation}
\label{eq:lowerbound}
\underline{\delta}(\cdot)=\frac{\sum_{v\in V_{\min}} f(v,q)}{k}
\end{equation}

\vspace{0.1cm}
\begin{myLemma}
\label{lemma:2}
Given an arbitrary substate $\tilde{H}^j_k\subseteq \tilde{H}^i_k$, its attribute distance must be lower bounded by $\underline{\delta}(\cdot)$, i.e., $\delta(\tilde{H}^j_k)\geq \underline{\delta}(\cdot)$.
\end{myLemma}

\vspace{0.1cm}
This naturally holds as any $\tilde{H}^j_k$ containing at least one node with $f(\cdot,q)\geq \max\{f(v,q)|v\in V_{\min}\}$ would increase $\delta(\tilde{H}^j_k)$.

\vspace{0.1cm}
\noindent \underline{(2) Prune based on $\underline{\delta}(\cdot)$.} For each visited state $\tilde{H}^i_k$, we record its $\delta(\tilde{H}^i_k)$ to update the optimal $\delta^*(\cdot)$ so far, compute the lower bound $\underline{\delta}(\cdot)$, and decide to prune according to Theorem \ref{th:4}.

\vspace{0.1cm}
\begin{myTheorem}
\label{th:4}
Given a state $\tilde{H}^i_k$, we say it is unpromising to find a better state $\tilde{H}^j_k\subseteq \tilde{H}^i_k$ with a smaller $\delta(\tilde{H}^j_k)$ than the optimal $\delta^*(\cdot)$ and can be safely pruned, if $\underline{\delta}(\cdot)\geq \delta^*(\cdot)$.
\end{myTheorem}

\vspace{0.1cm}
\begin{IEEEproof}
\label{proof:th4}
Given $\underline{\delta}(\cdot)\geq\delta^*(\cdot)$ and Lemma \ref{lemma:2}, we have $\delta(\tilde{H}^j_k)\geq \underline{\delta}(\cdot)\geq \delta^*(\cdot)$. So, we can prune $\tilde{H}^i_k$ safely.
\end{IEEEproof}

\vspace{0.1cm}
\begin{myExample}
\label{example:3}
In Figure \ref{fig:prune}, suppose we are now at state $\tilde{H}^3_2$ and the current optimal community is formed by nodes $\{v_3,v_5,v_6\}$ with $\delta^*(\cdot)=0.45$. The lower bound $\underline{\delta}(\cdot)$ for all substates of $\tilde{H}^3_2$ is $\frac{0.6+0.5}{2}$ $=$ $0.55$ $>$ $\delta^*(\cdot)$, so we can prune $\tilde{H}^3_2$ safely.
\end{myExample}

\setlength{\textfloatsep}{0cm}
\begin{algorithm}[t]
\setstretch{0.85}
\fontsize{9.5pt}{4mm}\selectfont
\caption{Enumeration with three prunings}
\label{alg:whole_alg}
\LinesNumbered
\KwIn{The maximal connect $k$-core $\tilde{H}_k$, a query node $q$}
\KwOut{The connect $k$-core $\tilde{H}^*_k$ with the smallest $\delta(\cdot)$}
$\tilde{H}^*_k \leftarrow \tilde{H}_k$, $\delta^*(\cdot) \leftarrow \delta(\tilde{H}_k)$\;
$u\leftarrow {\sf NULL}$ \tcc*{{\scriptsize Previously deleted node}}
$f(u,q)\leftarrow +\infty$\;
Enumerate($\tilde{H}_k$, $q$, $u$)\;
\Return $\tilde{H}^*_k$\;
\nonl \quad\\
\nonl \textbf{Procedure} Enumerate($\tilde{H}_k$, $q$, $u$)\\
\tcp{{\footnotesize Prune for unpromising states}}
$\underline{\delta}(\cdot) \leftarrow$ lower bound for substates $\subseteq \tilde{H}_k$ \tcc*{{\scriptsize Eq. \ref{eq:min_k1}-\ref{eq:lowerbound}}}
\If{$\underline{\delta}(\cdot)\geq\delta^*(\cdot)$} 
{
	\Return \tcc*{{\scriptsize Theorem \ref{th:4}}}
}
\Else{
	\tcp{{\footnotesize Prune for unnecessary states}}
	$D_{>\delta} \leftarrow$ nodes of $f(\cdot,q)>\delta(\tilde{H}_k)$ \tcc*{{\scriptsize Theorem \ref{th:3}}}
	\While{$D_{>\delta}\neq \emptyset$}
	{
		$v \leftarrow D_{>\delta}$.pop\_max() \tcc*{{\scriptsize node to delete}}
		$\langle\tilde{H}^i_k,v_m\rangle\leftarrow$ maintain a $k$-core for $\tilde{H}_k\setminus v$\;
		\tcp{{\footnotesize Prune for duplicated states}}
		\If{$f(v_m,q)>f(u,q)$}
		{
			\Continue \tcc*{{\scriptsize Theorem \ref{th:2}}}
		}
		\Else{
			\If{$\delta(\tilde{H}^i_k)<\delta^*(\cdot)$}
			{
				$\tilde{H}^*_k \leftarrow \tilde{H}^i_k$, $\delta^*(\cdot) \leftarrow \delta(\tilde{H}^i_k)$\; 
			}
			\tcc{{\scriptsize Update previously deleted node}}
			$u\leftarrow v$\;
			Enumerate($\tilde{H}^i_k$, $q$, $u$)\;
		}
	}
	\Return\;
}
\end{algorithm}

\vspace{0.1cm}
\noindent \textbf{Combine three pruning strategies together.} Algorithm \ref{alg:whole_alg} shows the whole procedure of enumeration with three pruning strategies. We study the effect of prunings on \texttt{Exact}'s efficiency in \S \ref{sec:efficiency}. Given the maximal connected $k$-core $\tilde{H}_k$ and a query node $q$, we first initialize the optimal community $\tilde{H}^*_k$ and attribute distance $\delta^*(\cdot)$ as $\tilde{H}_k$ and $\delta(\tilde{H}_k)$ (line 1). Since $\tilde{H}_k$ is the root of enumeration, it doesn't have a previously deleted node $u$ ($u$ is used to indicate that the current $\tilde{H}_k$ is obtained from its parent state by deleting $u$). Thus we configure $f(u,q)$ as positive infinity (lines 2-3) and use it in the \textit{prune for duplicated states}. We next enumerate from $\tilde{H}_k$ as follows: (1) We compute the lower bound of attribute distance $\underline{\delta}(\cdot)$ for all substates of $\tilde{H}_k$ via Eq. \ref{eq:min_k1}-\ref{eq:lowerbound}, and then \textit{prune all unpromising substates} if $\underline{\delta}(\cdot)>\delta^*(\cdot)$ (lines 6-8: Theorem \ref{th:4}). (2) For each $\tilde{H}_k$ that was not pruned before, we record all nodes with $f(\cdot,q)>\delta(\tilde{H}_k)$ from $\tilde{H}_k$ in a max-heap $D_{>\delta}$, and  \textit{prune all unnecessary substates} by deleting nodes from $D_{>\delta}$ (lines 10-11: Theorem \ref{th:3}). (3) Given a $D_{>\delta}\neq \emptyset$, we pop node $v$ with the largest $f(\cdot,q)$ and maintain a new connected $k$-core $\tilde{H}^i_k$ by deleting $v$ from $\tilde{H}_k$. During the $k$-core maintenance, we also record node $v_m$ with the largest $f(\cdot,q)$ that is recursively deleted after $v$ is removed (lines 12-13). If $f(v_m,q)>f(u,q)$, then we can \textit{prune this duplicated state} according to Theorem \ref{th:2} (lines 14-15). Otherwise, we update the optimal $\tilde{H}^*_k$ and $\delta^*(\cdot)$ if necessary (lines 17-18), update the previously deleted node $u$ as the current deleted node $v$ (line 19), keep enumerating from this new $\tilde{H}^i_k$ (line 20), and finally return the optimal community $\tilde{H}^*_k$ (line 5).



\subsection{Complexity Analysis}
\label{sec:complexity}
For the exact baseline without  pruning strategies, the time complexity in the worst case is $O(|E_G|+|V_{\tilde{H}_k}|!\cdot |E_{\tilde{H}_k}|)$. The first term $|E_G|$ is the time of finding the maximal connected $k$-core $\tilde{H}_k$ from the entire graph $G$ via core-decomposition \cite{batagelj2003m}. While the second term $|V_{\tilde{H}_k}|!\cdot |E_{\tilde{H}_k}|$ is the time complexity of enumeration over $\tilde{H}_k$, where $|V_{\tilde{H}_k}|!$ is the total number of states in the search tree and $|E_{\tilde{H}_k}|$ is the time for core-maintenance on each state. Since the first pruning strategy prunes the duplicated states, the second term can be reduced to $2^{|V_{\tilde{H}_k}|}\cdot |E_{\tilde{H}_k}|$. While the last two pruning strategies can further reduce the number of states, thus we introduce a constant $C$ ($\in [2,1000]$ in practice) to represent the time as $2^{|V_{\tilde{H}_k}|}/C\cdot |E_{\tilde{H}_k}|$.


\vspace{0.1cm}
\begin{myExample}
\label{exp:complexity}
In IMDB, considering the community search with $k=22$, the average \# states for 200 random queries when using the first pruning strategy is nearly $1.47\times 10^{11}$. After applying the last two pruning strategies, the \# states is reduced to $1.52\times 10^8$. In this case, we have $C=967.1$.
\end{myExample}

\section{Sampling-Estimation Solution}
\label{sec:sampling_solution}
We next present an \textit{sampling-estimation}-based approximate method to improve \texttt{Exact}'s efficiency from two aspects: reducing the size of the maximal $\tilde{H}_k$ based on \textit{sampling} and terminating the enumeration early if a reliable accuracy (in the form of a confidence interval) is obtained based on \textit{estimation}. 

\begin{figure}
\setlength{\abovecaptionskip}{0.1cm}
\vspace{-0.2cm}
  \centering
  \includegraphics[scale=0.44]{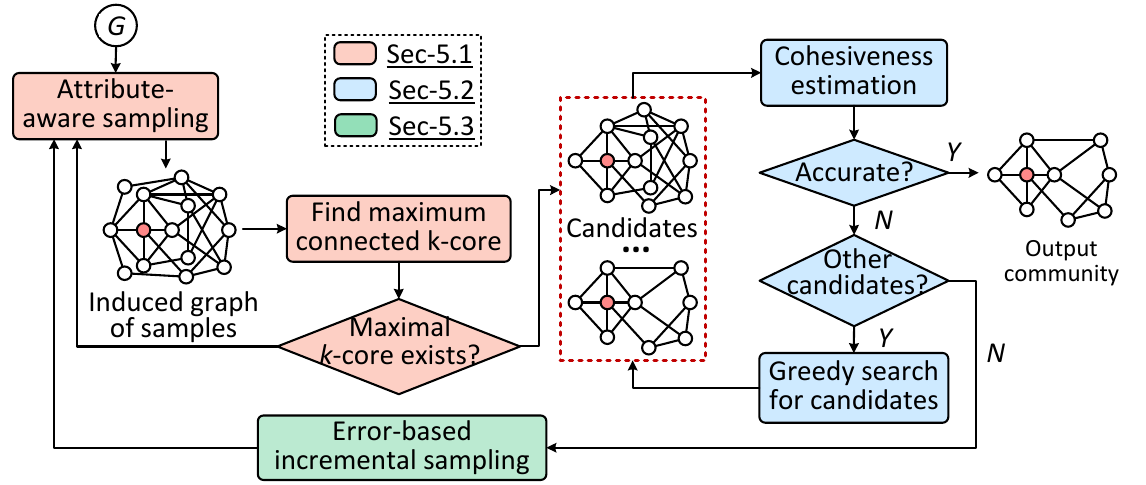}
  \caption{The pipeline of our sampling-estimation method}
  \label{fig:pipeline}
\end{figure}

Figure \ref{fig:pipeline} shows the pipeline with three steps: (1) \textit{Sampling-based maximal $\tilde{H}_k$ finding} (\textbf{\S \ref{sec:sampling}}). We determine the minimum population for sampling through \textit{Hoeffding Inequality} \cite{Hoeffding1994}, collect a set of samples (nodes) $S$ via an attribute-aware sampling, and take the maximal $\tilde{H}_k$ from the induced graph of $S$ as the input for estimation. (2) \textit{Estimation with accuracy guarantee} (\textbf{\S \ref{sec:acc_guarantee}}). We estimate the $\delta(\cdot)$ of each candidate community through \textit{Bag of Little Bootstrap} \cite{Kleiner2014} and terminate early when an accurate enough result is obtained. Otherwise, we iteratively find and estimate another candidate by \textit{greedy search}. (3) \textit{Error-based incremental sampling} (\textbf{\S \ref{sec:error_sampling}}). If we cannot find a satisfactory community, then we enlarge $S$ via an error-based incremental sampling and repeat steps 1-2.


\subsection{Sampling-based Maximal $\tilde{H}_k$ Finding}
\label{sec:sampling}
A straightforward idea is to collect a set of nodes (as samples $S$) that are similar to the query node $q$ from the entire graph $G$. Given a node $v\in V_G$, we can compute $v$'s sampling probability $P_s(v)$ as a normalized value (Eq. \ref{eq:prob}). The smaller the attribute distance $f(v,q)$, the greater the $P_s(v)$.
\begin{equation}
\label{eq:prob}
P_s(v)=\frac{1-f(v,q)}{\sum_{u\in V_G}(1-f(u,q))}
\end{equation}


The biggest issue is that the sampling population (i.e., $G$) is too large to guarantee the sample quality. According to small world theory \cite{Kleinberg2000,Malkov2020}, two nodes in the same cohesive community exhibit strong access locality \cite{Yang2012}. This implies that nodes from the neighborhood of $q$ are more likely to belong to the community of $q$. With this in mind, we prefer to collect samples from $q$'s neighborhood (denoted by $G_q\subseteq G$) instead of the entire $G$. Obviously, the size of $G_q$ (i.e., \# nodes in $G_q$) is important, a large (small) $G_q$ may contain more (less) irrelevant (relevant) nodes w.r.t. $q$, thereby affecting the sample quality \cite{Cheng2023,Wang2022,Wang2014}. A critical problem is how to set an appropriate size of $G_q$ in order to bound the sample quality. Straightforwardly, we can simply define an $h$-hop $G_q$ of $q$. However, it's problematic, as an empirical $h$ cannot adapt various $k$ of $k$-core over different datasets. Hence, we resort to \textit{Hoeffding Inequality} to determine the minimum size of $G_q$. 

\vspace{0.1cm}
\noindent \textbf{Minimum size of $G_q$.} Given a graph $G$ and a query node $q$, we first introduce the existence probability of a node $u\in V_G$ that belongs to $q$'s neighborhood $G_q$, denoted by $P(u)$ (discussed later), which is computed based on the node's sampling probability $P_s(\cdot)$. The larger the $P(u)$, the higher the probability that $u$ is included in a specific $G_q$. Suppose the ground-truth community of CS-AG is $\tilde{H}_k$. Let $v\in V_{\tilde{H}_k}$ be the node with the smallest existence probability $P(v)$. \textit{We expect to find the minimum size of $G_q$ that every node $u\in V_{\tilde{H}_k}$ with $P(u)\geq P(v)$ would be included in $G_q$ with a large probability.} In this way, sampling from $G_q$ can be viewed as a good approximation to sampling from $G$, because $G_q$ includes sufficient relevant nodes w.r.t. $q$ (i.e., nodes from $\tilde{H}_k$ with $P(u)\geq P(v)$) and the least irrelevant nodes w.r.t. $q$ (i.e., minimizing $G_q$). In the following, we leverage \textit{possible world semantics} \cite{Bonchi2014,Miao2022,Peng2018} to compute a node's existence probability, then we apply Hoeffding Inequality to the existence probability to find the minimum size of $G_q$.

\vspace{0.1cm}
\noindent\underline{Possible worlds w.r.t. $G_q$.} Given a graph $G$ and a query node $q$, we may easy to obtain various neighborhoods of $q$ (e.g., different $G_q$ with different size) by collecting nodes based on their normalized sampling probabilities $P_s(v)$ (Eq. \ref{eq:prob}). According to possible world semantics, each $G_q$ can be seen as a possible world. Thus, we can measure an edge's existence probability $P(e)$ in a possible world as Eq. \ref{eq:edge_prob}, which implies that an edge $e_{uv}$ exists in a $G_q$ only if its two end nodes $u$ and $v$ are sampled simultaneously. 
\begin{equation}
\label{eq:edge_prob}
P(e_{uv})=P_s(u)\times P_s(v)
\end{equation}

Given the edge's existence probability above, we can measure a specific $G_q$'s existence probability as follows.
\begin{equation}
\label{eq:world_prob}
P(G_q)=\prod\limits_{e\in G_q} P(e)\prod\limits_{e\in G\setminus G_q}(1-P(e))
\end{equation}

Given the aforementioned analysis, we show the probability of a node $v$ that belongs to $q$'s neighborhood is equal to the aggregate probability over all possible worlds \cite{Cheng2023} (Eq. \ref{eq:world_prob}). $\mathcal{G}_q$ represents all possible worlds, and $I_{G_q}(v)$ is an indicator function denoting if $v$ belongs to $G_q$ ($I_{G_q}(v)=1$) or not.
\begin{equation}
\label{eq:world_prob}
P(v)=1-\prod\limits_{u\in N(v)}(1-P(e_{uv}))=\sum\limits_{G_q\in \mathcal{G}_q}P(G_q)\times I_{G_q}(v)
\end{equation}

\vspace{0.1cm}
\noindent\underline{Hoeffding Inequality.} We next leverage $P(v)$ to determine the minimum size of $G_q$, ensuring that all nodes of the ground-truth community are included in $G_q$ with a large probability.

\vspace{0.1cm}
\begin{myTheorem}
\label{th:hoeffding}
{\rm (Hoeffding Inequality \cite{Hoeffding1994} w.r.t. $P(v)$)} Given a set of possible worlds $\{G^1_q,\dots,G^t_q\}$ and an estimation error $\epsilon>0$, let $\hat{P}(v)$ be the estimation of $P(v)$, where $\hat{P}(v)=\sum P(G^i_q)\times I_{G^i_q}(v)$ and $\hat{P}(v)\in[a,b]$ ($a=0$,$b=1$). Then, we have the following inequality.
\begin{equation}
\label{eq:hoeffding}
{\rm Pr}[\hat{P}(v)-P(v)\geq \epsilon]\leq \exp(-\frac{2t^2\epsilon^2}{\sum_{i=1}^t(a-b)^2})
\end{equation}
\end{myTheorem}

According to Hoeffding Inequality, this theorem provides an upper bound on the probability of the fact that $\hat{P}(v)$ has a large estimation error to $P(v)$. It's obvious that the larger the $t$ (\# possible worlds), the smaller the upper bound it is, showing that $\hat{P}(v)-P(v)<\epsilon$ holds with a larger probability. Based on Theorem \ref{th:hoeffding}, we have the following theorem holds.

\vspace{0.1cm}
\begin{myTheorem}
\label{th:hoeffding1}
Given a set of possible worlds $\{G^1_q,\dots,G^t_q\}$, $\epsilon>0$, and $u$, $v\in V_G$, if $P(u)-P(v)\geq \epsilon$, then we have
\begin{equation}
\label{eq:hoeffding1}
{\rm Pr}[\hat{P}(v)-\hat{P}(u)>0]\leq \exp(-t\epsilon^2/2)	\quad .
\end{equation}
\end{myTheorem}

\begin{IEEEproof}
We consider $\hat{P}(v)-\hat{P}(u)$ as the estimator of $P(v)-P(u)$ with $\hat{P}(v)-\hat{P}(u)\in [-1,1]$. Then we have the following derivation by subjecting $\hat{P}(v)-\hat{P}(u)$ to Eq. \ref{eq:hoeffding}.
\end{IEEEproof}
\vspace{-0.3cm}
\begin{equation}\nonumber
\small
\begin{aligned}
{\rm Pr}[\hat{P}(v)-\hat{P}(u)>0] & \leq {\rm Pr}[\hat{P}(v)-\hat{P}(u)-(P(v)-P(u))\geq\epsilon]\\
& \leq \exp(-\frac{2t^2\epsilon^2}{\sum_{i=1}^t(1-(-1))^2})\\
& =\exp(-t\epsilon^2/2)
\end{aligned}
\end{equation}

Theorem \ref{th:hoeffding1} shows the theoretical result of bounding the order of a pair of nodes. That is, if we have $P(v)<P(u)$ for two nodes, then their estimated existence probabilities would satisfy $\hat{P}(v)\leq\hat{P}(u)$ with a probability $\geq 1-\exp(-t\epsilon^2/2)$.
More precisely, given the assumption of $P(v)\leq P(u)-\epsilon<P(u)$, the probability of $\hat{P}(v)>\hat{P}(u)$ is upper bounded by $\exp(-t\epsilon^2/2)$. The smaller the upper bound, the more the probability of $\hat{P}(v)\leq\hat{P}(u)$, showing that $u$ is more likely to be included in $G_q=\bigcup_{i=1}^t G^i_q$ than $v$, given $P(v)<P(u)$.  

Given the ground-truth community $\tilde{H}_k$ of CS-AG, let $v\in V_{\tilde{H}_k}$ be the node with the smallest existence probability $P(v)$ and there are $m$ nodes in $\tilde{H}_k$ with existence probabilities $\geq P(v)$. Then, we can use Theorem \ref{th:hoeffding1} to derive the minimum number of possible worlds (i.e., $t$) w.r.t. $G_q$ that ensures all $m$ nodes can be contained in $G_q$ with at least $1-\beta$ probability, as Theorem \ref{th:hoeffding2} shows. Specifically, Theorem \ref{th:hoeffding2} bounds the order of $m(n-m)$ pairs of nodes by applying \textit{Union Bound} and Theorem \ref{th:hoeffding1}, where $n=|V_G|$ is the number of nodes in $G$.

\vspace{0.1cm}
\begin{myTheorem}
\label{th:hoeffding2}
Given a desired probability of $1-\beta$ and $\epsilon>0$, it requires $t\geq \frac{2}{\epsilon^2}\ln \frac{m(n-m)}{\beta}$ possible worlds to ensure that $G_q$ contains all $m$ nodes (from $\tilde{H}_k$) with existence probability $\geq P(v)$, where $v$ is the node with the smallest existence probability in the ground-truth $\tilde{H}_k$.
\end{myTheorem}

\vspace{0.1cm}
\begin{IEEEproof}
Since we expect all $m$ nodes with $P(\cdot)\geq P(v)$ would be contained in $G_q$ with at least $1-\beta$ probability, we need to ensure that every node $u$ from such $m$ nodes has a larger $\hat{P}(u)$ than that of other $n-m$ nodes, i.e., we need to bound the order of $m(n-m)$ pairs of nodes. We have the following derivation through Bound Union and Theorem \ref{th:hoeffding1}.
\end{IEEEproof}
\vspace{-0.3cm}
\begin{equation}\nonumber
\begin{aligned}
1-\beta & \leq 1-m(n-m)\exp(-t\epsilon^2/2) \\
& \Rightarrow \beta\geq m(n-m)\exp(-t\epsilon^2/2) \Rightarrow t\geq \frac{2}{\epsilon^2}\ln \frac{m(n-m)}{\beta}\quad
\end{aligned}
\end{equation}

We next show how to compute the the minimum size of $G_q$ (Theorem \ref{th:hoeffding3}) based on the minimum $t$ given in Theorem \ref{th:hoeffding2}.

\vspace{0.1cm}
\begin{myTheorem}
\label{th:hoeffding3}
Given the ground-truth community $\tilde{H}_k$ of CS-AG, in the worst case, we require at least $\frac{2}{\epsilon^2}\ln \frac{(k+1)(n-k-1)}{\beta}+1$ nodes from the original $G$ to form $G_q$, so that all nodes in $\tilde{H}_k$ would be contained in $G_q$ with a probability of $1-\beta$.
\end{myTheorem}

\vspace{0.1cm}
\begin{IEEEproof}
A $k$-core has at least $k+1$ nodes, which means that we have at least $k+1$ nodes should be contained in $G_q$. Thus, we need to compare at least $(k+1)(n-k-1)$ pairs of nodes (i.e., $m=k+1$), which indicates that we need at least $\frac{2}{\epsilon^2}\ln \frac{(k+1)(n-k-1)}{\beta}$ possible worlds w.r.t. $G_q$ (Theorem \ref{th:hoeffding2}). In the worst case, each possible world w.r.t. $G_q$ can be an individual edge between $q$ and another node. Thus, we require at least $\frac{2}{\epsilon^2}\ln \frac{(k+1)(n-k-1)}{\beta}+1$ nodes to form the final $G_q$.
\end{IEEEproof}

\vspace{0.1cm}
\begin{myExample}
Given the DBLP with $682819$ nodes, $k=30$, $\epsilon=0.05$, and $1-\beta=98\%$, it requires at least $\frac{2}{0.05^2}\ln \frac{(31)(682819-31)}{0.02}+1$ $\approx$ $16625$ nodes to form a $G_q$.
\end{myExample}

\vspace{0.1cm}
We next conduct a BFS starting from the query node $q$ to form $G_q$. In this BFS, we preferentially expand the search from those nodes having smaller composite attribute distances to $q$, until the minimum size of $G_q$ is reached. In \S \ref{sec:parameter}, we investigate the effect of $\epsilon$ and $1-\beta$ on CS's performance.

\vspace{0.1cm}
\noindent \textbf{Attribute-aware sampling over $G_q$.} We perform attribute-aware sampling over the population $G_q$ as follows. We compute the sampling probabilities $P_s(\cdot)$ of all nodes in $G_q$ based on their composite attribute distances to $q$ by replacing $G$ by $G_q$ in Eq. \ref{eq:prob}. We randomly collect $|S|$ samples (i.e., nodes) from $G_q$ according to their $P_s(\cdot)$. We initialize the sample size as a fraction $\lambda$ of nodes in $G_q$, i.e., $|S|=\lambda\cdot |V_{G_q}|$, and update $|S|$ with an appropriate $|\Delta S|$ if necessary (discussed in \S \ref{sec:error_sampling}). In \S \ref{sec:parameter}, we show the parameter sensitivity of $\lambda$.

\vspace{0.1cm}
\noindent \textbf{Find the maximal connected $k$-core.} We maintain a maximal connected $k$-core from the induced graph $G_q[S]$ of samples $S$ and take it as input of the estimation step (\S \ref{sec:acc_guarantee}) to find an approximate community of $q$ for Approx-CS-AG problem.

\subsection{Estimation with Accuracy Guarantee}
\label{sec:acc_guarantee}
Given the maximal $\tilde{H}_k\subseteq G_q[S]$, a user-input error bound $e$, and a confidence level $1-\alpha$, Approx-CS-AG problem aims to find an approximate community $\tilde{H}^\star_k$ with an attribute distance $\delta^\star$ satisfying $|\delta^\star-\delta|/\delta\leq e$ with a probability of $1-\alpha$ (Eq. \ref{eq:CI}-\ref{eq:error_rate}). (1) We provide a confidence interval CI $=\delta^\star\pm \varepsilon$ at $1-\alpha$ level to quantify the quality of $\delta^\star$ based on \textit{Central Limit Theorem} (CLT) \cite{Wang2022, Wang2014}, and apply \textit{Bag of Little Bootstrap} (BLB) to compute the Margin of Error (MoE) $\varepsilon$ of CI. (2) We return $\tilde{H}^\star_k$ when a tight CI (i.e., $\varepsilon\leq \frac{\delta^\star\cdot e}{1+e}$, Theorem \ref{th:relative_error}) is obtained. (3) Otherwise, we greedily remove the most dissimilar node from $\tilde{H}^\star_k$ to get a new candidate $\tilde{H}^i_k\subseteq \tilde{H}^\star_k$ and repeat above. If we cannot find a good $\tilde{H}^\star_k$, we repeat steps (1)-(3) with enlarged samples $S=S\cup\Delta S$ (\S \ref{sec:error_sampling}).

\vspace{0.1cm}
\noindent\textbf{Confidence interval calculation.} Recall the attribute distance $\delta^\star$ is the average composite attribute distance $f(\cdot,q)$ of all nodes from $V_{\tilde{H}^\star_k}\setminus q$, where $f(v,q)$ of $\forall v\in V_{\tilde{H}^\star_k}$ is considered as a random variable with the sampling probability of $P_s(v)$. From CLT, we know that a mean-like point estimator follows a normal distribution \cite{Gao2019}. So, we have $\delta^{\star}$ $\sim$ $N(\mu_{\delta^\star},\sigma^2_{\delta^\star})$, and the MoE $\varepsilon$ of CI $=\delta^\star\pm \varepsilon$ at $1-\alpha$ level can be calculated based on CLT as $\varepsilon=z_{\alpha/2}\cdot \sigma_{\delta^{\star}}$, where $z_{\alpha/2}$ is the normal critical value with right-tail probability $\alpha/2$ (obtained from a standard normal table). We next use BLB to estimate $\sigma_{\delta^\star}$.

\vspace{0.1cm}
\noindent\underline{Bag of little bootstrap.} Bootstrap \cite{Efron1993} provides an automatic and widely applicable means of quantifying estimator quality \cite{Kleiner2012}. Though it's simple and powerful, it requires computing the estimators on resamples having a size comparable to the original data. If the original data are large, then bootstrap is costly. Thus, we resort to BLB \cite{Kleiner2012}, which incorporates features of both \textit{bootstrap} and \textit{subsampling}, for high-quality estimation with a quite small sample. Given the approximate community $\tilde{H}^\star_k$, we do BLB as follows: (1) We collect $s$ small subsamples $\{S_1,\dots, S_s\}$ from $V_{\tilde{H}^\star_k}$, $\forall S_i$ has a size of $|V_{\tilde{H}_k^\star}|^m$, where $m\in [0.5,1)$ is the scale factor used in \cite{Kleiner2012} to ensure $s\cdot |V_{\tilde{H}_k^\star}|^m\leq |V_{\tilde{H}_k^\star}|$. We use $S_{\rm blb}=\bigcup S_i$ to indicate all subsamples for BLB estimation. (2) For each $S_i$, BLB estimates $\sigma_{\delta^\star}$ by a standard bootstrap (given below) and computes an MoE $\varepsilon_i=z_{\alpha/2}\cdot \sigma_{\delta^{\star}}$. (3) Given $s$ MoEs $\{\varepsilon_1,\dots,\varepsilon_s\}$, BLB computes the final $\varepsilon=\sum \varepsilon_i/s$.

\vspace{0.1cm}
\noindent\underline{Bootstrap.} Given a subsample $S_i$, a standard bootstrap first collects $r$ resamples having size $|S_i|$ from $S_i$ with replacement. Then, it computes $\delta^\star$ for each resample as $\{\delta^\star_1,\dots, \delta^\star_r\}$. Next, it takes the empirical distribution of $\{\delta^\star_1,\dots, \delta^\star_r\}$ as an approximation to $N(\mu_{\delta^\star},\sigma^2_{\delta^\star})$, so we estimate $\sigma_{\delta^\star}$ by Eq. \ref{eq:blb}.
\begin{equation}
\label{eq:blb}
\mu_{\delta^\star}=\sum \delta^\star_i/r\ ,\quad \sigma_{\delta^\star}=\sum(\delta^\star_i-\mu_{\delta^\star})/(r-1)
\end{equation}


\noindent\textbf{Accuracy guarantee.} Given a CI = $\delta^\star\pm \varepsilon$, we ensure that the relative error of $\delta^\star$ is bounded by a user-input error bound $e$. 

\vspace{0.1cm}
\begin{myTheorem}
\label{th:relative_error}
If the MoE $\varepsilon$ satisfies $\varepsilon\leq \frac{\delta^\star\cdot e}{1+e}$, then the relative error is upper bounded by $e$ with a probability of $1-\alpha$.
\end{myTheorem}

\vspace{0.1cm}
\begin{IEEEproof}
We prove this theorem in the following two steps.

\noindent\textbf{Step 1.} 
Suppose that the exact $\delta$ locates in the CI's right half-width, i.e., $\delta^\star\leq \delta\leq \delta^\star+\varepsilon$. Then we have the following derivation and $(\delta-\delta^\star)/\delta\leq e$ holds if $\varepsilon/\delta^\star\leq e$ (i.e., $\varepsilon\leq\delta^\star\cdot e$).
\begin{equation}\nonumber
(\delta-\delta^\star)/\delta\leq (\delta-\delta^\star)/\delta^\star\leq \varepsilon/\delta^\star
\end{equation}

\noindent\textbf{Step 2.} 
Suppose that the exact $\delta$ locates in the CI's left half-width, i.e., $\delta^\star-\varepsilon\leq \delta\leq \delta^\star$. Then $(\delta^\star-\delta)/\delta\leq e$ holds if $\varepsilon/(\delta^\star-\varepsilon)\leq e$ (i.e., $\varepsilon\leq \frac{\delta^\star\cdot e}{1+e}$) by the following derivation.
\begin{equation}\nonumber
(\delta^\star-\delta)/\delta\leq \varepsilon/\delta\leq \varepsilon/(\delta^\star-\varepsilon)
\end{equation}

In summary, $|\delta^\star-\delta|/\delta\leq e$ holds if $\varepsilon\leq \frac{\delta^\star\cdot e}{1+e}$, as $\frac{\delta^\star\cdot e}{1+e}$ is a tighter bound ($\leq \delta^\star\cdot e$). Since our CI of $\delta^\star$ has a confidence level $1-\alpha$, the above holds with a probability of $1-\alpha$.
\end{IEEEproof}

\vspace{0.1cm}
\noindent\textbf{Greedy search of candidate communities.} If the accuracy guarantee (Theorem \ref{th:relative_error}) is not satisfied, we keep enumerating from $\tilde{H}^\star_k$ to get another candidate $\tilde{H}^i_k\subseteq \tilde{H}^\star_k$ for further estimation. A straightforward method is to apply our \textit{enumeration with prunings} (\S \ref{sec:enumeration}) to enumerate a new candidate and do BLB estimation until Theorem \ref{th:relative_error} holds. Given the premise of finding an approximate community, we can simplify it to a greedy candidate search (without backtracking) by deleting the most dissimilar node at each state. Specifically, given a current $\tilde{H}^i_k$, we delete node $v$ with the most dissimilar composite attribute distance $f(v,q)$ to $q$, then we maintain the maximal $\tilde{H}_k$ of the remaining nodes as the next candidate and do BLB estimation for it. We terminate it once Theorem \ref{th:relative_error} holds. 

\vspace{-0.1cm}
\subsection{Error-based Incremental Sampling}
\label{sec:error_sampling}
If we cannot find a good $\tilde{H}^\star_k$, then we should enlarge the samples $S=S\cup\Delta S$. Intuitively, we need a large $|\Delta S|$ when the MoE $\varepsilon$ is large. Otherwise, a small $|\Delta S|$ is sufficient. So, we present a method to automatically set $|\Delta S|$ based on $\varepsilon$.

Consider an MoE $\varepsilon>\frac{\delta^\star\cdot e}{1+e}$. We use $\varepsilon/\frac{\delta^\star\cdot e}{1+e}$ to denote how far $\varepsilon$ is away from the desired value $\frac{\delta^\star\cdot e}{1+e}$. The larger $\varepsilon/\frac{\delta^\star\cdot e}{1+e}$ is, the more nodes that $\Delta S$ requires. Ideally, if we can reduce $\varepsilon$ to a new $\varepsilon'$ by at least $\varepsilon/\frac{\delta^\star\cdot e}{1+e}$ times, we can satisfy $\varepsilon'\leq \frac{\delta^\star\cdot e}{1+e}$. Since $\varepsilon=z_{\alpha/2}\cdot \sigma_{\delta^{\star}}$, reducing $\varepsilon$ by $\varepsilon/\frac{\delta^\star\cdot e}{1+e}$ times is equivalent to reducing $\sigma_{\delta^\star}$ by $\varepsilon/\frac{\delta^\star\cdot e}{1+e}$ times. Since $\sigma_{\delta^\star}=\sigma_{\delta}/\sqrt{|V_{\tilde{H}_k}|}$ according to CLT, where $\sigma_{\delta}$ is the standard deviation of the population, we say that reducing $\sigma_{\delta^\star}$ by $\varepsilon/\frac{\delta^\star\cdot e}{1+e}$ times is equivalent to increasing $|V_{\tilde{H}_k}|$ by $(\varepsilon/\frac{\delta^\star\cdot e}{1+e})^2$ times. In summary, we can increase $|V_{\tilde{H}_k}|$ by $(\varepsilon/\frac{\delta^\star\cdot e}{1+e})^2$ times to reduce $\varepsilon$ by $\varepsilon/\frac{\delta^\star\cdot e}{1+e}$ times. Hence, we derive $|\Delta S|$ as follows. 
\vspace{-0.15cm}
\begin{equation}
\label{eq:delta}
|\Delta S|=|S_{\rm blb}|\cdot [(\varepsilon/\frac{\delta^\star\cdot e}{1+e})^{2m}-1]
\end{equation}
\vspace{-0.15cm}
%


\vspace{-0.1cm}
\begin{myExample}
\label{exp:deltaS}
Given a CI = $\delta^\star\pm \varepsilon$ with $\delta^\star$ = $0.3$, $\varepsilon$ = $3.5\times 10^{-3}$, and $|S_{\rm blb}|$ = $1000$. If we set the scale factor $m$ = $0.6$ and error bound $e$ = $0.01$, then we need $|\Delta S|$ = $1000\cdot ((3.5\times 10^{-3}/\frac{0.3\cdot 0.01}{1.01})^{2\cdot 0.6}-1)\approx 253$ to update $S$. While for a large $\varepsilon=8\times 10^{-3}$, we then require $|\Delta S|\approx 2284$ to update $S$.
\end{myExample}

\subsection{Complexity Analysis}
\label{sec:complexity1}
The total time of our approximate solution consists of sampling time ($T_s$) and estimation time ($T_e$). For sampling, we fist require a BFS to get a neighbrood graph $G_q$ of $q$ with a time of $|V_{G_q}|+|E_{G_q}|$, then we collect $|S|=\lambda\cdot |V_{G_q}|$ samples from $G_q$ and find the maximal connected $k$-core $\tilde{H}_k$ from the induced graph $G_q[S]$ with a time of $|S|+|V_{G_q[S]}|$. So, we get $T_e=O((1+\lambda)\cdot |V_{G_q}|+|E_{G_q}|+|V_{G_q[S]}|)$. For estimation, we introduce a constant $N_e$ to indicate the number of iterations of accuracy estimation till termination condition (Theorem \ref{th:relative_error}) is reached ($N_e\leq 5$ in practice). In each iteration, we greedily enumerate from $\tilde{H}_k\subseteq G_q[S]$ by deleting one node with the most dissimilar attribute distance, for each explored state we perform BLB estimation over $|S_{\rm blb}|$ subsamples. So, the total time of BLB estimation is $|V_{\tilde{H}_k}|\cdot |S_{\rm blb}|$. If we cannot find a good community, then we should include $|\Delta S|$ additional samples for the next iteration of accuracy estimation. Thus, the total time is $T_e=O(N_e\cdot (|V_{\tilde{H}_k}|\cdot |S_{\rm blb}|+|\Delta S|))$.

\section{Extensions}
\label{sec:extension}
We extend our sampling-estimation solution to three more general scenarios: (1) CS on heterogeneous graphs, (2) Size-bounded CS, and (3) CS with different community models.
 
\subsection{Extension to Heterogeneous Graphs}
\label{sec:hetero}
A heterogeneous graph $G$ consists of a node (edge) set $V_G$ ($E_G$) with multiple node types $\mathcal{T}$ ($\mathcal{R}$). For $\forall v\in V_G$, it has a node type $\phi(v):V_G\rightarrow \mathcal{T}$. For $\forall e\in E_G$, it has an edge type $\psi(e):E_G\rightarrow \mathcal{R}$. The meta-path $\mathcal{P}$ is often used to indicate a specific relationship between two nodes with the same type. For example, in DBLP, $\mathcal{P}=\texttt{A}$-$\texttt{P}$-$\texttt{A}$ shows the co-authorship w.r.t. a paper between two authors. We call nodes with the type linked by $\mathcal{P}$ as target nodes (e.g., authors for $\texttt{A}$-$\texttt{P}$-$\texttt{A}$) and we aim to find an approximate $(k,\mathcal{P})$-core community \cite{Fang2020,Wang2022,Wang2022a} of target nodes from $G$ satisfying the constrains of Approx-CS-AG problem. We refer readers to \cite{Fang2020,Xu2022,Wang2022a} for more details of heterogeneous graphs and meta-path. We extend our method to this scenario with three modifications: (1) We replace \# nodes $n=|V_G|$ in Theorem \ref{th:hoeffding3} with \# target nodes of $G$ to compute the minimum size of $G_q$. (2) We construct $G_q$ by a $\mathcal{P}$-neighbor-oriented BFS from the query node $q$, which expands the search by exploring a node's $\mathcal{P}$-neighbors. We say two target nodes are $\mathcal{P}$-neighbors if they are connected by a path instance of $\mathcal{P}$. (3) We perform BLB estimation on the community of target nodes, using the attribute distance $\delta(\cdot)$ computed by target nodes' $f(\cdot,q)$ to $q$. 

Besides the basic $(k,\mathcal{P})$-core, there are several variants of $(k,\mathcal{P})$-core. For example, (1) $(k,\mathcal{P})$-truss \cite{Yang2020} is an extension of $(k,\mathcal{P})$-core, we can support it by the same method as above but change the core-maintenance to truss-maintenance during the BLB estimation. (3) Heterogeneous influential community (HIC) is proposed in \cite{Zhou2023}, it aims to identify a $(k,\mathcal{P})$-core community $H$ satisfying that there is no other community $H'$ with the influence vector $f(H')$ dominates the influence vector $f(H)$ of $H$. The dominance relationship is defined the same as skyline, that is for each element $f_i(H')$ in $f(H')$, it must $\geq f_i(H)$. We may support HIC with a modification on the BLB estimation, i.e., estimating the MAX value of each element in the influence vector of an approximate community $H$. More precisely, we may resort to Extreme Value Theory (EVT) \cite{Coles2001} to conduct EVT-based MAX value estimation \cite{Wang2022} for each element in the influence vector of an approximate community.


\subsection{Extension to Size-bounded CS}
\label{sec:size}
The community's size is critical to some applications \cite{Yao2021,Ma2019}. Many applications naturally require that the number of members in a community should fall within a certain range, e.g., organize a workshop with at least $l$ attendees and no more than $h$ attendees. This motivates a size-bounded CS \cite{Yao2021} to find a community with size $\in [l,h]$. We extend our sampling-estimation solution to size-bounded CS with three modifications: (1) We require at least $\frac{2}{\epsilon^2}\ln \frac{l(n-l)}{\beta}+1$ nodes to construct $G_q$, because the desired community' size is lower-bounded by $l$. Thus, we replace $k+1$ in Theorem \ref{th:hoeffding3} with $l$ to get the new minimum size of $G_q$. (2) We ignore the candidate communities with size $>h$ during the estimation and stop the greedy search of candidates when we have a size $\leq l$. (3) We early terminate the estimation when we get a community having the size $\in [l,h]$ and the MoE $\varepsilon\leq \frac{\delta^\star\cdot e}{1+e}$ (Theorem \ref{th:relative_error}). 

\begin{figure*}
  \setlength{\abovecaptionskip}{-0.1cm}
  \setlength{\belowcaptionskip}{0cm}
  \hspace{-0.7cm}
  \subfigure[Attribute distance $\delta$]{
  \begin{minipage}{0.23\linewidth}
  \includegraphics[scale=0.42]{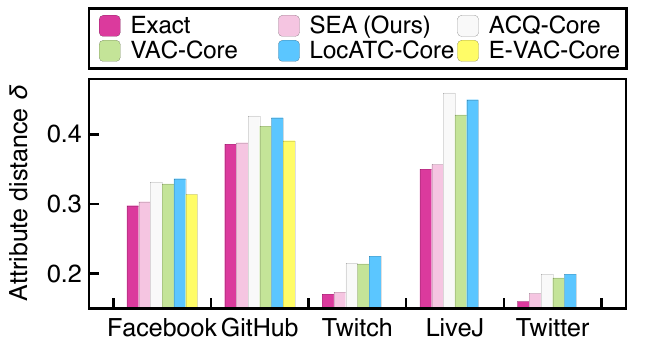}
  \vspace{-0.3cm}
  \end{minipage}
  }
  \hspace{0.2cm}
  \subfigure[Relative error (\%) of $\delta$]{
  \begin{minipage}{0.23\linewidth}
    \includegraphics[scale=0.42]{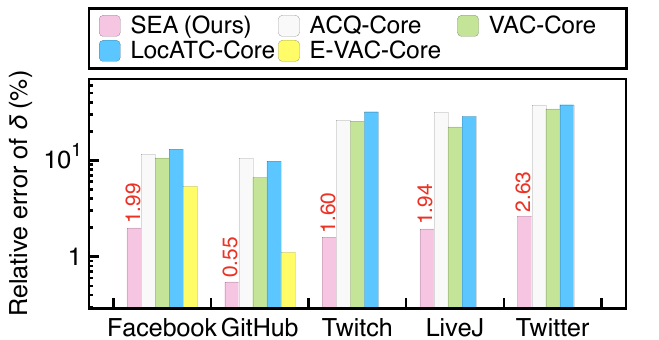}
    \vspace{-0.3cm}
  \end{minipage}
  }
 \hspace{0.2cm}
  \subfigure[Response time (ms)]{
  \begin{minipage}{0.23\linewidth}
    \includegraphics[scale=0.42]{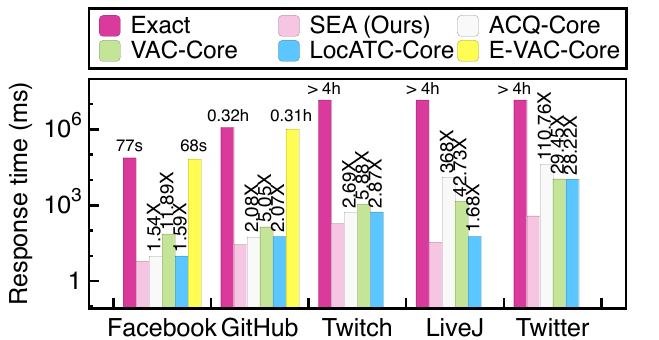}
    \vspace{-0.3cm}
  \end{minipage}
  }
  \hspace{0.2cm}
  \subfigure[Response time (ms) of three steps]{
  \begin{minipage}{0.23\linewidth}
    \includegraphics[scale=0.42]{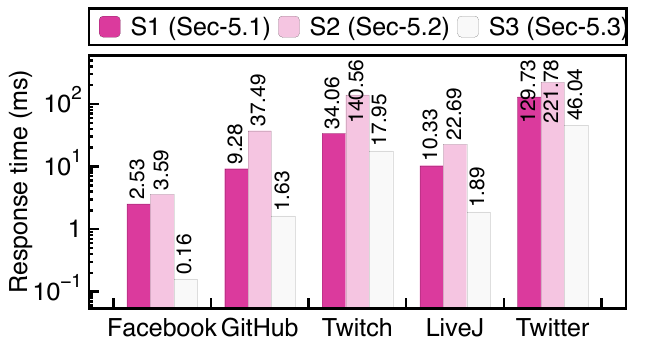}
    \vspace{-0.3cm}
  \end{minipage}
  }
  \caption{Effectiveness (a)-(b) and efficiency (c)-(d) results over homogeneous graphs}
  \label{fig:effective_efficiency1}
  \vspace{-0.5cm}
\end{figure*}

\subsection{Extension to Various Community Models}
\label{sec:models}
Besides $k$-core, $k$-truss \cite{Huang2015} is another popular model to measure a community's structure cohesiveness. According to \cite{fang2020survey}, it is widely-recognized that $k$-truss shows a higher structure cohesiveness but is less efficient than $k$-core. Users may choose an appropriate model based on their actual demands. We extend our sampling-estimation solution to be adaptive to $k$-truss model with three modifications: (1) For a $k$-truss, every node must have at least $k-1$ neighbors, indicating it is a $(k$-1$)$-core. So, it has at least $k$ nodes and we update the minimum size of $G_q$ as $\frac{2}{\epsilon^2}\ln \frac{k(n-k)}{\beta}+1$ (Theorem \ref{th:hoeffding3}). (2) Given the induced graph $G_q[S]$ of $S$, we find the maximal connected $k$-truss from it instead of connected $k$-core, as input of BLB estimation. (3) During the estimation, we maintain a connected $k$-truss as a candidate community instead of $k$-core.

\setlength{\textfloatsep}{0.1cm}
\begin{table}
\setlength{\abovecaptionskip}{0.1cm}
\centering
\caption{Statistics of datasets}
\label{tab:datasets}
\scalebox{0.6}{
\hspace{-0.8cm}
\begin{tabular}{c||c|c|c|c|c|c|c|c}
\textbf{Datasets} $\downarrow$ & \#\textbf{Nodes} & \#\textbf{Edges}  & \#\textbf{N-types} & \#\textbf{E-types} & $\bm d_{\rm max}$ & $\bm d_{\rm avg}$ & $\bm k_{\rm max}$ & $\bm k_{\rm avg}$ \\ \hline \hline
Facebook & 4,039 & 88,234 & 1 & 1 & 1,045 & 43.69 & 117 & 22.44  \\ \hline
GitHub		& 37,700 & 289,003  & 1 & 1 & 9,458 & 15.33 & 36 & 7.12   \\ \hline
Twitch		& 168,114  & 6,797,557 & 1 & 1 & 35,259  & 80.86  & 151  & 36.72    \\ \hline
LiveJournal & 3,997,962 & 34,681,189 & 1 & 1 & 14,815 & 17.34 & 362 & 7.84 \\ \hline 
Twitter-2010		&	21,297,772	&	265,025,810	&	1	&	1	&	698,112 & 24.88  & 1,695 & 12.71  	\\ \hline
DBLP 		&	682,819	&	1,951,209 &	 4	&	6	& 345 & 3.75 & 28 & 2.64 	\\ \hline
IMDB		&	2,875,685	&	9,705,602	&	4	&	24 & 591 & 4.42 & 552 & 4.37  \\ \hline
DBpedia  & 4,521,912 & 15,045,801 & 359 & 676 & 6760 & 289.79 & 422 & 149  \\ \hline
Freebase		&	5,706,539	&	48,724,743	&	11,666	&	5,118	& 467	 & 5.64 & 60 & 2.75 \\ \hline
YAGO 	& 7,308,072 & 36,624,106 & 6543 & 101 & 285 & 5.31 & 44 & 2.61   \\
\end{tabular} 
}
\end{table}

\section{Experiments}
\label{sec:exp}
We provide the experimental study on ten real-world datasets. Our code \cite{code} was implemented in Java 1.8 and run on a 2.1 GHZ, 64 GB memory AMD-6272 Linux server. Our evaluation seeks to answer the following questions.

\noindent \textbf{Q1:} How do our \texttt{Exact} and approximate solutions (\S \ref{sec:exact}-\ref{sec:sampling_solution}) perform in effectiveness and efficiency? (\textbf{\S \ref{sec:effective}-\ref{sec:efficiency}})

\noindent \textbf{Q2:} What is the effect of pruning strategies? (\textbf{\S \ref{sec:prune}})

\noindent \textbf{Q3:} How is the scalability of approximate method? (\textbf{\S \ref{sec:scale}})

\noindent \textbf{Q4:} How does our method find an approximate community iteratively on real-world datasets? (a case study in \textbf{\S \ref{sec:case}})

\noindent \textbf{Q5:} How do parameters (discussed in \S \ref{sec:sampling_solution}) affect the approximate method's effectiveness and efficiency? (\textbf{\S \ref{sec:parameter}})

\vspace{-0.05cm}
\subsection{Experimental Setup}
\label{sec:setup}
\noindent\textbf{Datasets.} Table \ref{tab:datasets} summarizes some statistics, e.g., the maximum (average) coreness $k_{\rm max}$ ($k_{\rm avg}$) and degree $d_{\rm max}$ ($d_{\rm avg}$), of 5 homogeneous and 5 heterogeneous graphs. (1) Facebook \cite{McAuley2012}, (2) GitHub \cite{Rozemberczki2021}, (3) Twitch \cite{Rozemberczki2021a}, (4) LiveJournal \cite{Yang2012}, and (5) Twitter-2010 \cite{Rossi2015} are social networks. (6) DBLP \cite{dblp} provides relationships among authors, papers, venues, etc. Each author has several attributes, e.g., research interests, \# publications, $h$-index and \# citations. (7) IMDB \cite{imdb} provides relationships among actors, directors, and movies, with attributes like category, \# movies for actors; genres, ratings for movies. (8) DBpedia \cite{DBpedia}, (9) Freebase \cite{Bollacker2008}, and (10) YAGO \cite{Rebele2016} are well-known knowledge graphs. Similar to \cite{Wang2022}, we add attributes for several types of nodes via web crawling.


\vspace{0.1cm}
\noindent\textbf{Queries.} We generated 200 queries for each graph. For homogeneous graphs, we follow \cite{Fang2016} to form a query with a random query node. For heterogeneous graphs, we generate a query the same as \cite{Fang2020}. We first obtain the top-10 meta-paths with the highest frequencies. A meta-path $\mathcal{P}$'s frequency is measured by its \# path instances. The more the path instances, the higher the frequency of $\mathcal{P}$. We next form a query with a randomly selected $\mathcal{P}$ and a query node with the type linked by $\mathcal{P}$. 

\vspace{0.1cm}
\noindent\textbf{Metrics.} We use the attribute distance $\delta(\cdot)$, relative error of $\delta(\cdot)$ w.r.t. the ground-truth (obtained by \texttt{Exact}) to evaluate the effectiveness. We evaluate the efficiency by response time. We show the average result of 200 queries in each test.

\vspace{0.1cm}
\noindent\textbf{Methods.} We implemented \texttt{Exact} and \textbf{S}ampling-\textbf{E}stimation-based \textbf{A}pproximate solution (SEA) for $k$-core (default) and $k$-truss: (1) \texttt{Exact}, (2) \texttt{Exact}-Truss, (3) SEA, and (4) SEA-Truss. We compare ours with representative CS methods using various attribute cohesiveness metrics: (5) LocATC-Core and (6) LocATC-Truss, the fastest local version of ATC \cite{Huang2017} atop $k$-core and $k$-truss, which are two approximate methods. (7) ACQ-Core \cite{Fang2016} is an exact core-based method. (8) VAC-Core is the core-based version of the truss-based (9) VAC-Truss \cite{Liu2020}, both two are approximate methods. (10) E-VAC-Core and (11) E-VAC-Truss are the corresponding exact VAC methods also from \cite{Liu2020}. Ours (1)-(4) support two types of graphs, while (5)-(11) are designed for homogeneous graphs. We convert a graph from heterogeneous to homogeneous given a meta-path, then invoke (5)-(11) to find communities for heterogeneous graphs. Besides, we clarified that we only provide the results of E-VAC-Core for small graphs, i.e., Facebook and GitHub, because it cannot finish within one week for large graphs \cite{Liu2020}.


\vspace{0.1cm}
\noindent\textbf{Parameters.} The default parameters are: $k\geq 4$, $1-\beta=95\%$ and $\epsilon=0.05$ for Hoeffding Inequality, $\lambda=0.2$ is the initial sampling fraction, $1-\alpha=95\%$ and $e=0.02$ for accuracy guarantee. We show the parameter sensitivity in \S \ref{sec:parameter}.

\vspace{0.1cm}
\noindent\textbf{Remark.} Since some datasets provide human-annotated ground-truth (HA-GT) community, e.g., Facebook, LiveJournal, Orkut \cite{Orkut}, and Amazon \cite{Yang2015}, we also evaluate the effectiveness w.r.t. HA-GT using $F_1$-score as the metric (the same as \cite{Huang2017}). The higher the $F_1$-score, the more the similarity of a community to the HA-GT, showing that the community with strict structure and attribute cohesiveness constraints can reflect the characteristics of real communities to some extent.

\setlength{\textfloatsep}{0.2cm}
\begin{table}
\setlength{\abovecaptionskip}{0cm}
\centering
\caption{Various attribute cohesiveness ({\small Facebook)}}
\label{tab:attri_cohesiveness}
\scalebox{0.65}{
\begin{tabular}{c||c|c|c|c||c}
\textbf{Methods} $\downarrow$ & \makecell[c]{Min-max \\ (VAC)} & \makecell[c]{Attribute \\ coverage (ATC)} &  \makecell[c]{\#Shared \\ attributes (ACQ)} & $\delta(\cdot)$ (Ours) & Total rank \\ \hline \hline
SEA (Ours)	& \underline{0.486} (2)	& 161.84 (4) & \underline{0.06} (2) & \underline{0.304} (2) & \textbf{10}  \\ \hline
LocATC-Core	& 0.491 (6)  & \textbf{209.39} (1) & 0.04 (6) & 0.331 (6) & 19  \\ \hline
ACQ-Core	&	0.489 (5)  & \underline{196.79} (2) & \textbf{0.08} (1) & 0.328 (5) & 13 \\ \hline
VAC-Core	& \underline{0.486} (2) & 178.46 (3)  & \underline{0.06} (2) & 0.325 (4)  & \underline{11}\\ \hline
\texttt{Exact} (Ours) 	& \underline{0.486} (2) 	& 155.13 (6) & \underline{0.06} (2) & \textbf{0.297} (1) & \underline{11} \\ \hline
E-VAC-Core	& \textbf{0.475} (1) 	& 158.45 (5) & \underline{0.06} (2) & 0.314 (3) & \underline{11} \\
\end{tabular} 
}
\vspace{-0.2cm}
\end{table}

\subsection{Effectiveness Evaluation}
\label{sec:effective}
Figure \ref{fig:effective_efficiency1}(a) shows the results of homogeneous graphs. SEA has smaller $\delta(\cdot)$ than others and it is quite close to that of \texttt{Exact}. From the perspective of relative error of $\delta(\cdot)$ (Figure \ref{fig:effective_efficiency1}(b)), \textit{ours is at least one order of magnitude less than others and is bounded by the default error bound $e=2\%$}. This is because we apply BLB estimation with an reliable accuracy guarantee (Theorem \ref{th:relative_error}). Moreover, we measure each method's attribute cohesiveness w.r.t. various metrics. Table \ref{tab:attri_cohesiveness} shows the results on Facebook (associated with a rank in parentheses). We highlight the best results in bold and indicate suboptimal values with underlines. Each method performs the best on its own metric. From the macro perspective (see total rank), SEA is the best for all metrics. We also apply the same method as \cite{Huang2017} to evaluate the $F_1$-score w.r.t. HA-GT community (Table \ref{tab:hagt}). Here, we use `-' to indicate that a method cannot finish within a sufficiently long time. SEA and \texttt{Exact} have higher $F_1$-score than others, indicating that the community of ours is more similar to HA-GT than others. We also provide the $F_1$-score over 10 ego-networks of Facebook in Figure \ref{fig:facebook} and we found that ours has the best $F_1$-score on eight of them.


\setlength{\textfloatsep}{0.1cm}
\begin{table}
\setlength{\abovecaptionskip}{0cm}
\centering
\caption{$F_1$-score w.r.t. HA-GT}
\label{tab:hagt}
\scalebox{0.71}{
\begin{tabular}{c||c|c|c|c}
\textbf{Methods} $\downarrow$ & \makecell[c]{Facebook} & \makecell[c]{LiveJournal} &  \makecell[c]{Orkut} & \makecell[c]{Amazon} \\ \hline \hline
SEA (Ours)	&  \underline{0.61}	& \underline{0.86} & \underline{0.56} & \underline{0.91} \\ \hline
LocATC-Core	&  0.54	&  0.76 & 0.45  & 0.73  \\ \hline
ACQ-Core	&  0.31	&  0.31  & 0.28 & 0.45  \\ \hline
VAC-Core	&  0.47	&  0.79 & 0.40  &  0.76 \\  \hline
\texttt{Exact} (Ours) 	& \textbf{0.64} & \textbf{0.88} & \textbf{0.58} & \textbf{0.92} \\ \hline
E-VAC-Core 	& 0.51 & - & - & - \\
\end{tabular} 
}
\vspace{0cm}
\end{table}

\begin{figure}[t]
\setlength{\abovecaptionskip}{0.1cm}
\vspace{-0.2cm}
  \centering
  \includegraphics[scale=0.51]{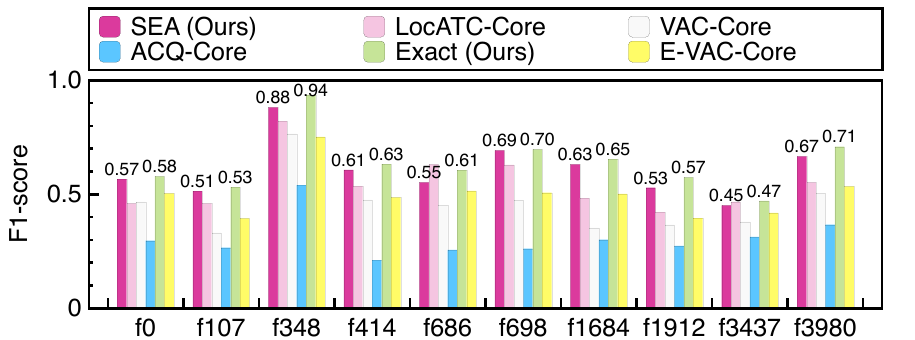}
  \caption{$F_1$-score on each ego-network provided in Facebook}
  \label{fig:facebook}
\vspace{-0.1cm}
\end{figure}

\subsection{Efficiency Evaluation}
\label{sec:efficiency}
Figure \ref{fig:effective_efficiency1}(c) shows the results of homogeneous graphs. We provide the speedup of SEA w.r.t. comparing methods, on the top of bars. Ours outperforms others and the improvement is getting obvious as the graph size increases. For ten-million-scale Twitter, \textit{ours is at least 28.2$\times$ faster than others}. For all datasets, \textit{ours is at least 1.54$\times$ (41.1$\times$ on average) faster than others}. This is because our method can early terminate once an acceptable $\delta(\cdot)$ is obtained. Figure \ref{fig:effective_efficiency1}(d) shows the runtime of SEA's three steps. S1: Sampling-based maximal $\tilde{H}_k$ finding (\S \ref{sec:sampling}). S2: BLB estimation (\S \ref{sec:acc_guarantee}). S3: Error-based incremental sampling (\S \ref{sec:error_sampling}). S2 is the most time-consuming step, as a greedy search is used to find candidate communities for BLB estimation. S3 is the most efficient step because in most cases, we can find a good community within 2 iterations.

\begin{table}[t]
\setlength{\abovecaptionskip}{0cm}
\caption{Effect of prunings on \texttt{Exact}'s efficiency (runtime in seconds and explored \# states during the enumeration)}
\label{tab:effect_prune}
\hspace{-0.4cm}
\scalebox{0.55}{
\begin{tabular}{c||cc||cc||cc||cc}
\multirow{2}{*}{\textbf{Methods} $\downarrow$}      & \multicolumn{2}{c||}{Facebook}           & \multicolumn{2}{c||}{GitHub}             & \multicolumn{2}{c||}{Twitch}             & \multicolumn{2}{c}{LiveJournal}     \\
      & \multicolumn{1}{c|}{Time} & \# States & \multicolumn{1}{c|}{Time} & \# States & \multicolumn{1}{c|}{Time} & \# States & \multicolumn{1}{c|}{Time} & \# States \\ \hline\hline
\texttt{Exact} & \multicolumn{1}{c|}{77}      &  3.24$\times 10^6$  & \multicolumn{1}{c|}{1210}      & 8.02$\times 10^7$   & \multicolumn{1}{c|}{14721}      & 1.07$\times 10^9$  & \multicolumn{1}{c|}{59292}      & 4.13$\times 10^9$  \\ \hline
\texttt{Exact}$\setminus$P3   & \multicolumn{1}{c|}{80}     & 3.35$\times 10^6$   & \multicolumn{1}{c|}{1890}     & 1.24$\times 10^8$    & \multicolumn{1}{c|}{15770}      & 1.16$\times 10^{9}$    & \multicolumn{1}{c|}{59315}      & 4.29$\times 10^{9}$   \\ \hline
\texttt{Exact}$\setminus$P3+P2   & \multicolumn{1}{c|}{388}     & 8.23$\times 10^7$   & \multicolumn{1}{c|}{174483}     & 4.21$\times 10^{10}$    & \multicolumn{1}{c|}{753701}     & 5.48$\times 10^{10}$    & \multicolumn{1}{c|}{$>$8 days}      & 1.02$\times 10^{12}$    \\ \hline
\texttt{Exact} w/o P & \multicolumn{1}{c|}{$>$8 days}      & 6.87$\times 10^{10}$   & \multicolumn{1}{c|}{$>$8 days}     & 8.79$\times 10^{12}$   & \multicolumn{1}{c|}{$>$8 days}     & 2.81$\times 10^{12}$ & \multicolumn{1}{c|}{$>$8 days}     & 4.51$\times 10^{15}$    \\
\end{tabular}
}
\vspace{-0.1cm}
\end{table}

\vspace{-0.1cm}
\subsection{Effect of Pruning Strategies for Exact Method}
\label{sec:prune}
In \S \ref{sec:exact}, we propose \texttt{Exact} with three pruning strategies to prune duplicated states (P1), unnecessary states (P2), and unpromising states (P3). Table \ref{tab:effect_prune} shows the effect of P1-P3 on the efficiency. \texttt{Exact} is the one with P1-P3, $\texttt{Exact}\setminus$P3 is the one with P1+P2, $\texttt{Exact}\setminus$P3+P2 is the one with P1, and $\texttt{Exact}$ w/o P is the one without prunings. Note that, all strategies are effective and improve the runtime. Among them, P1 is the most efficient one which can significantly reduce \# states explored in the searching. For example, for Facebook, P1 prunes 99.8\% states comparing with \texttt{Exact} w/o P.

\begin{table}[t]
\vspace{-0.1cm}
\setlength{\abovecaptionskip}{0cm}
\caption{Response time (ms) and relative error of $\delta$ (\%) for core- and truss-based methods on heterogeneous graphs.}
\label{tab:extension1}
\hspace{-0.4cm}
\scalebox{0.58}{
\begin{tabular}{c||cc||cc||cc||cc||cc}
\multirow{2}{*}{\textbf{Methods} $\downarrow$}      & \multicolumn{2}{c||}{DBLP}           & \multicolumn{2}{c||}{IMDB}             & \multicolumn{2}{c||}{DBpedia}             & \multicolumn{2}{c||}{Yago}              & \multicolumn{2}{c}{Freebase}          \\
      & \multicolumn{1}{c|}{Time} & Error & \multicolumn{1}{c|}{Time} & Error & \multicolumn{1}{c|}{Time} & Error & \multicolumn{1}{c|}{Time} & Error & \multicolumn{1}{c|}{Time} & Error \\ \hline\hline
SEA (Ours) & \multicolumn{1}{c|}{\textbf{187.01}}      & \textbf{1.58}   & \multicolumn{1}{c|}{\textbf{72.89}}      & \textbf{1.56}   & \multicolumn{1}{c|}{\textbf{59.64}}      & \textbf{0.0082}   & \multicolumn{1}{c|}{\textbf{76.57}}      & \textbf{1.26}  & \multicolumn{1}{c|}{\textbf{51.97}}      &  \textbf{1.43}  \\ \hline
ACQ-Core    & \multicolumn{1}{c|}{799.34}     & 13.45   & \multicolumn{1}{c|}{850.26}     & 41.57    & \multicolumn{1}{c|}{-}      & -    & \multicolumn{1}{c|}{-}      & - & \multicolumn{1}{c|}{-}      &  -  \\ \hline
LocATC-Core   & \multicolumn{1}{c|}{431.84}     & 14.58   & \multicolumn{1}{c|}{891.54}     & 47.83    & \multicolumn{1}{c|}{102.85}     & 37.58    & \multicolumn{1}{c|}{178.57}      & 20.10  & \multicolumn{1}{c|}{109.82}      &  24.81  \\ \hline
VAC-Core & \multicolumn{1}{c|}{1453.82}      & 12.45   & \multicolumn{1}{c|}{2700.96}     & 23.87   & \multicolumn{1}{c|}{397.70}     & 25.58 & \multicolumn{1}{c|}{562.72}     & 18.99  & \multicolumn{1}{c|}{447.73}      &  19.28  \\ \hline\hline
SEA-Truss & \multicolumn{1}{c|}{\textbf{334.57}}      & \textbf{0.21}   & \multicolumn{1}{c|}{\textbf{89.67}}      & \textbf{1.15}   & \multicolumn{1}{c|}{\textbf{72.99}}      & \textbf{1.23}   & \multicolumn{1}{c|}{\textbf{93.59}}      & \textbf{1.17}  & \multicolumn{1}{c|}{\textbf{64.26}}      &  \textbf{1.81}  \\ \hline
LocATC-Truss   & \multicolumn{1}{c|}{812.93}     & 4.89   & \multicolumn{1}{c|}{947.19}     & 21.29    & \multicolumn{1}{c|}{211.28}     & 19.99  & \multicolumn{1}{c|}{297.14}      & 15.28  & \multicolumn{1}{c|}{191.34}      &  15.17  \\ \hline
VAC-Truss & \multicolumn{1}{c|}{1857.71}      & 6.25   & \multicolumn{1}{c|}{2938.27}     & 9.04   & \multicolumn{1}{c|}{791.85}     & 2.37 & \multicolumn{1}{c|}{839.47}     & 5.31  & \multicolumn{1}{c|}{621.54}      &  5.87  \\
\end{tabular}
}
\end{table}

\subsection{Scalability Analysis}
\label{sec:scale}
\noindent\textbf{Extension to heterogeneous graphs}. We provide the results of core-based methods on five heterogeneous graphs in Table \ref{tab:extension1} (rows 1-4). Since we only have numerical attributes for DBpedia, Yago, and Freebase, the equality-matching-based method ACQ-Core cannot return any communities that share at least one numerical attribute. For all datasets, \textit{ours has at least one order of magnitude less relative error than others} and is bounded by the default $e=2\%$ (Theorem \ref{th:relative_error}). Besides, \textit{ours is at least 1.72$\times$ faster than others}, as we can terminate early when Theorem \ref{th:relative_error} holds.

\vspace{0.1cm}
\noindent\textbf{Extension to $k$-truss model}. Table \ref{tab:extension1} (rows 5-7) show the results for truss-based methods on heterogeneous graphs. Ours outperforms others due to the aforementioned same reasons.

\begin{figure}[h]
\centering
  \vspace{-0.3cm}
  \setlength{\abovecaptionskip}{0cm}
  \setlength{\belowcaptionskip}{0cm}
  \subfigure[DBLP]{
  \begin{minipage}{0.44\linewidth}
    \includegraphics[scale=0.35]{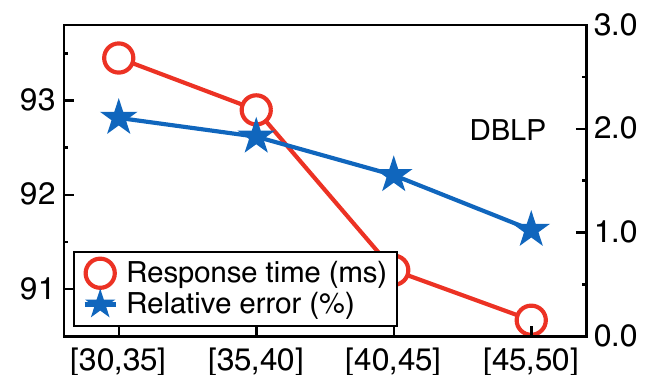}
    \vspace{-0.2cm}
  \end{minipage}
  }
  \hspace{0.25cm}
  \subfigure[GitHub]{
  \begin{minipage}{0.44\linewidth}
    \includegraphics[scale=0.35]{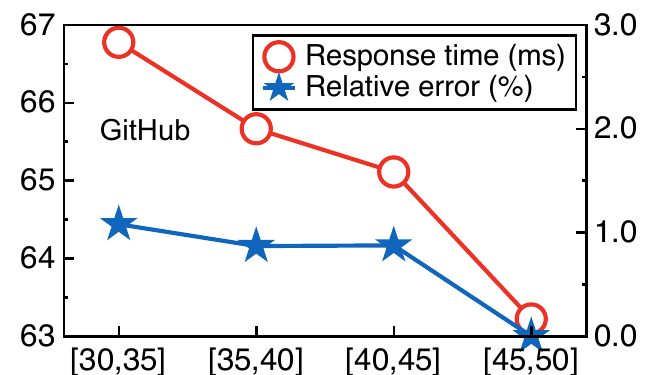}
    \vspace{-0.2cm}
  \end{minipage}
  }
  \caption{Results of size-bounded CS (SEA)}
  \label{fig:size_bound}
  \vspace{-0.2cm}
\end{figure}

\vspace{0.1cm}
\noindent\textbf{Extension to size-bounded CS}. Figure \ref{fig:size_bound} shows the results of size-bounded CS on DBLP and GitHub. The runtime decreases as the size increases, because the larger the community is desired, the less the time is required for greedy search of candidate communities. Besides, the relative error is bounded by the default $e=2\%$, showing BLB estimation is effective.

\begin{figure*}
  \setlength{\abovecaptionskip}{-0.1cm}
  \setlength{\belowcaptionskip}{0cm}
  \subfigure[DBLP-$\lambda$]{
  \begin{minipage}{0.13\linewidth}
    \includegraphics[scale=0.27]{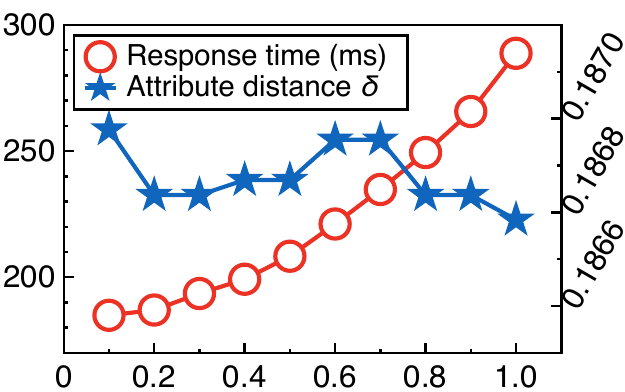}
    \vspace{-0.35cm}
  \end{minipage}
  }
  \hspace{0.2cm}
  \subfigure[Twitter-$\lambda$]{
  \begin{minipage}{0.13\linewidth}
    \includegraphics[scale=0.27]{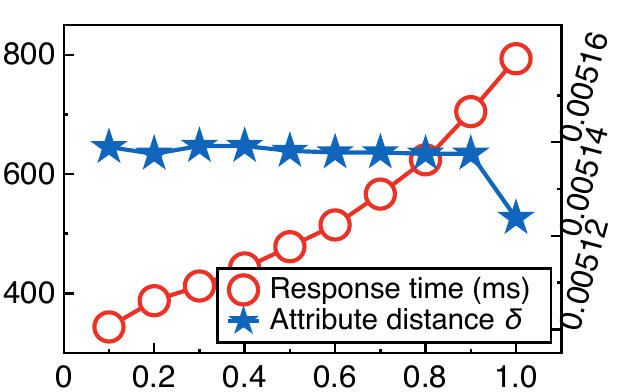}
    \vspace{-0.35cm}
  \end{minipage}
  }
  \hspace{0.2cm}
  \subfigure[DBLP-$\epsilon$]{
  \begin{minipage}{0.13\linewidth}
    \includegraphics[scale=0.27]{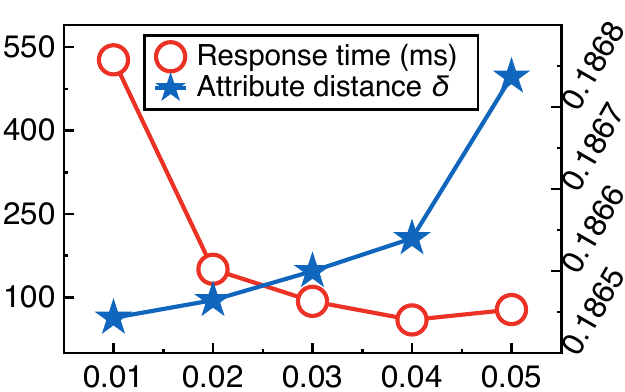}
    \vspace{-0.35cm}
  \end{minipage}
  }
  \hspace{0.2cm}
  \subfigure[Twitter-$\epsilon$]{
  \begin{minipage}{0.13\linewidth}
    \includegraphics[scale=0.27]{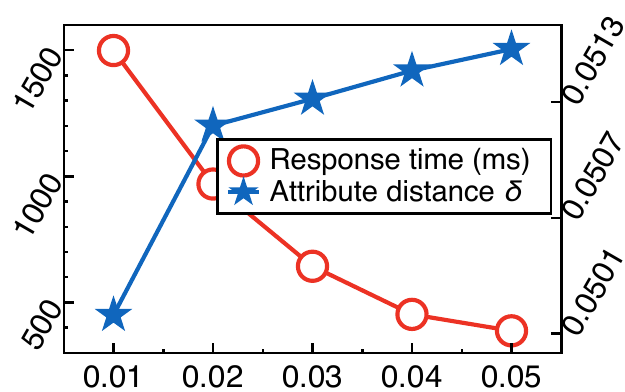}
    \vspace{-0.35cm}
  \end{minipage}
  }
  \hspace{0.2cm}
   \subfigure[DBLP-(1-$\beta$)]{
  \begin{minipage}{0.13\linewidth}
    \includegraphics[scale=0.27]{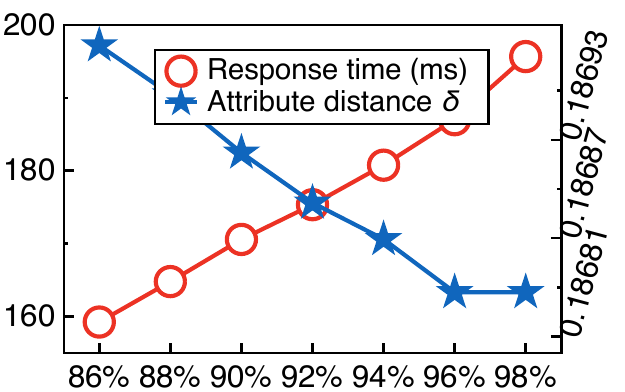}
    \vspace{-0.35cm}
  \end{minipage}
  }
  \hspace{0.2cm}
  \subfigure[Twitter-(1-$\beta$)]{
  \begin{minipage}{0.13\linewidth}
    \includegraphics[scale=0.27]{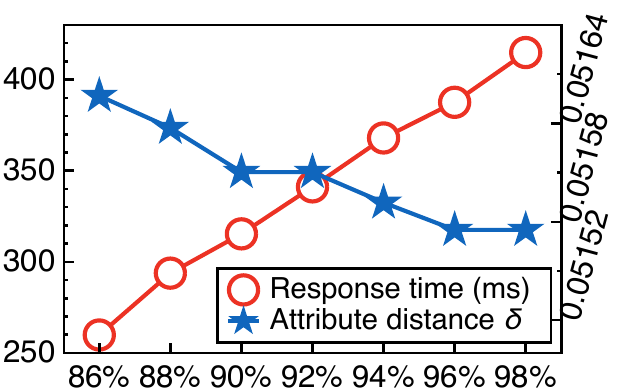}
    \vspace{-0.35cm}
  \end{minipage}
  }
  
  \vspace{-0.35cm}
  \subfigure[DBLP-$e$]{
  \begin{minipage}{0.13\linewidth}
    \includegraphics[scale=0.27]{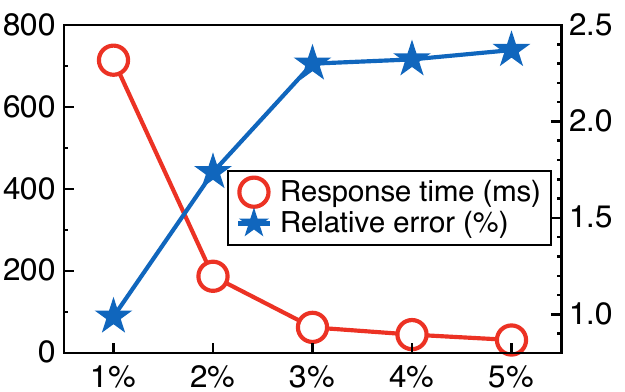}
    \vspace{-0.35cm}
  \end{minipage}
  }
  \hspace{0.2cm}
   \subfigure[Twitter-$e$]{
  \begin{minipage}{0.13\linewidth}
    \includegraphics[scale=0.27]{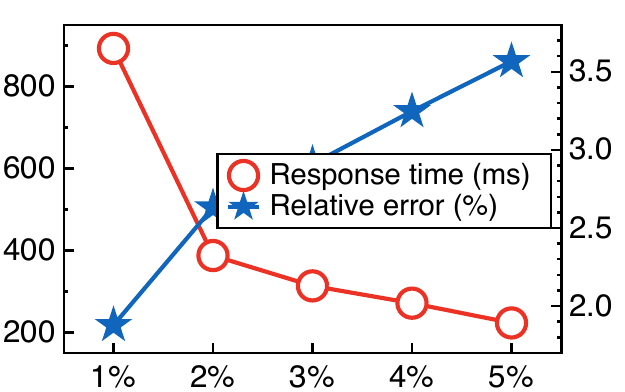}
    \vspace{-0.35cm}
  \end{minipage}
  }
  \hspace{0.2cm}
   \subfigure[DBLP-(1-$\alpha$)]{
  \begin{minipage}{0.13\linewidth}
    \includegraphics[scale=0.27]{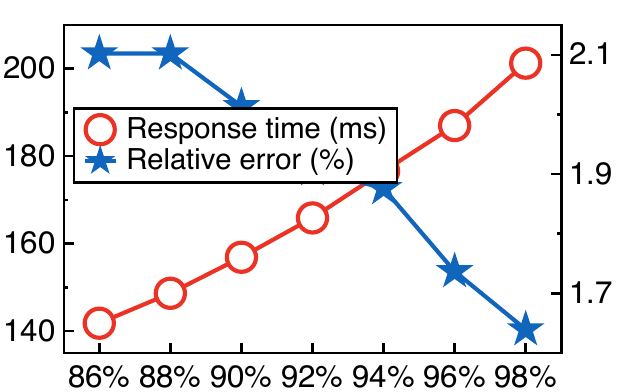}
    \vspace{-0.35cm}
  \end{minipage}
  }
  \hspace{0.2cm}
  \subfigure[Twitter-(1-$\alpha$)]{
  \begin{minipage}{0.13\linewidth}
    \includegraphics[scale=0.27]{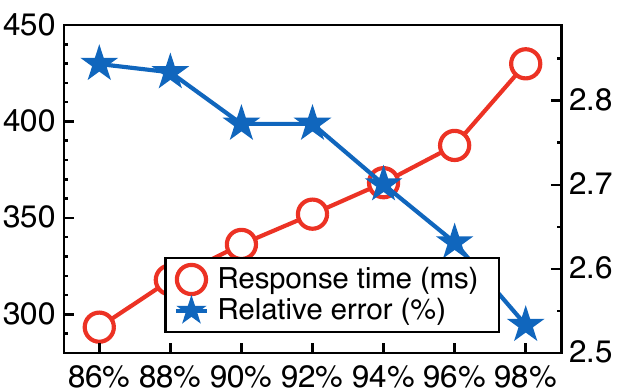}
    \vspace{-0.35cm}
  \end{minipage}
  }
  \hspace{0.2cm}
  \subfigure[DBLP-$k$]{
  \begin{minipage}{0.13\linewidth}
    \includegraphics[scale=0.27]{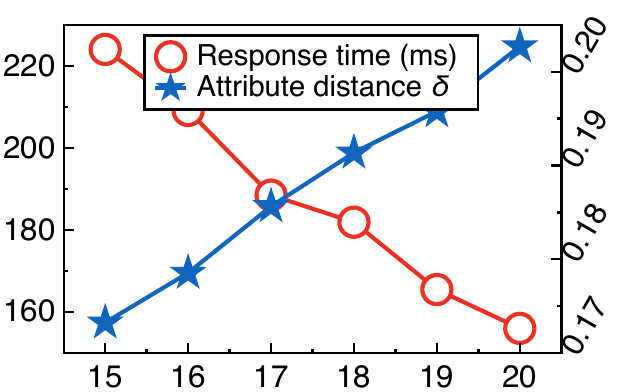}
    \vspace{-0.35cm}
  \end{minipage}
  }
  \hspace{0.2cm}
   \subfigure[Twitter-$k$]{
  \begin{minipage}{0.13\linewidth}
    \includegraphics[scale=0.27]{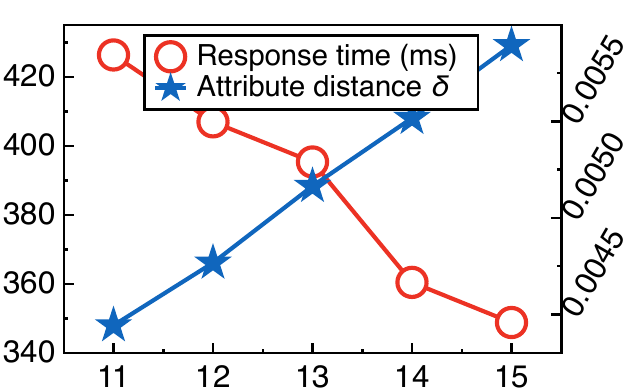}
    \vspace{-0.35cm}
  \end{minipage}
  }
  \caption{Parameter sensitivity (DBLP and Tiwtter): efficiency (left Y axis) and effectiveness (right Y axis).}
  \label{fig:parameter}
  \vspace{-0.6cm}
\end{figure*}

\begin{figure}[t]
\setlength{\abovecaptionskip}{-0.1cm}
\vspace{-0.4cm}
  \centering
  \includegraphics[scale=0.42]{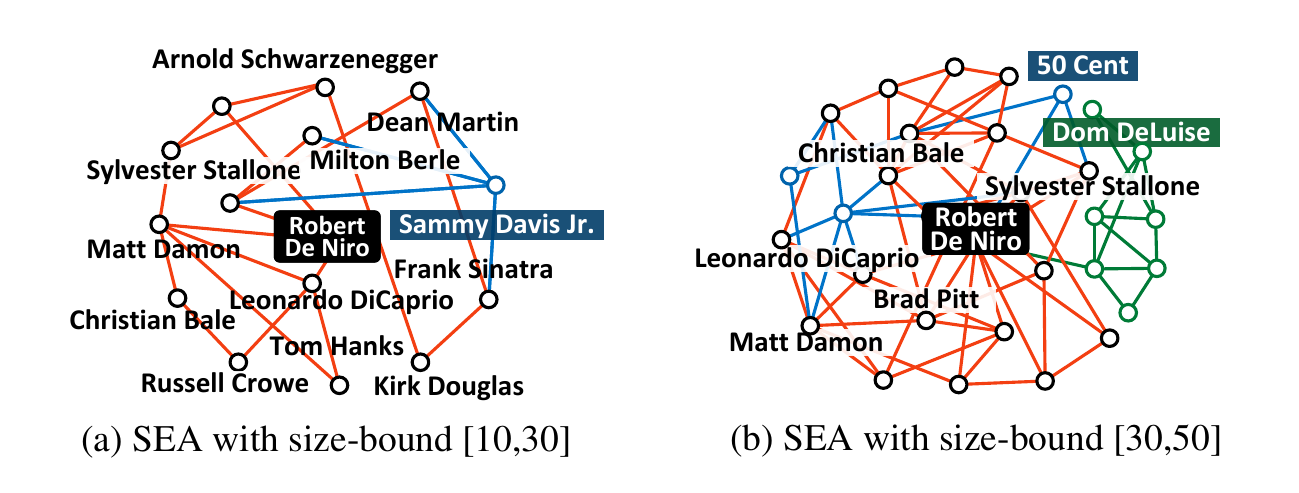}
  \caption{Case study of SEA with different size-bound}
  \label{fig:case_study}
\vspace{-0.2cm}
\end{figure}


\begin{table}[t]
\setlength{\abovecaptionskip}{0cm}
\setlength{\belowcaptionskip}{-0.1cm}
  \caption{Case study: detailed runtime information of SEA}
  \label{tab:case_study}
\scalebox{0.6}{
  \centering
\begin{tabular}{c||c|c|c|c|c|c}
\multirow{2}{*}{\textbf{Methods} $\downarrow$}  & \multicolumn{6}{c}{Approximate result round by round} \\ 
                                                                                & Round   &      $\delta^\star$       &  MoE $\varepsilon$    & $\Delta S$    & Time (ms)     & \tabincell{c}{Error (\%)}  \\ \hline
\multirow{2}{*}{SEA w/ size-bound $[10,30]$}            &     1   &   4.39$\times 10^{-3}$    &   9.23$\times 10^{-5}$     &   5967    &  48.29    & 2.34 \\ 
                                                                                &    2    &    4.31$\times 10^{-3}$   &    1.79$\times 10^{-5}$     &   615    &   3.85   & 0.39 \\ \hline
\multirow{2}{*}{SEA w/ size-bound $[30,50]$}            &     1   &   5.43$\times 10^{-3}$    &   1.48$\times 10^{-4}$     &   6743    &  52.45    & 4.66 \\ 
                                                                                &    2    &    5.17$\times 10^{-3}$   &    1.05$\times 10^{-4}$     &   3989    &   18.84   & 0.41 \\
\end{tabular}
}
\end{table}

\vspace{-0.2cm}
\subsection{Case Study}
\label{sec:case}
We performed a case study on IMDB with $q=$ Robert De Niro. Figure \ref{fig:case_study} illustrates two communities returned by SEA with different size-bound $[10,30]$ and $[30,50]$. We use different colors to distinguish persons with different levels of attribute similarities to Robert De Niro (Red $\succeq$ Blue $\succeq$ Green). The community in (a) includes a set of top-tire actors in Hollywood who are as famous as Robert De Niro, showing a greater attribute cohesiveness w.r.t. $q$ than that of (b). While (b) has to includes few less similar persons, e.g., 50 Cent and Dom Deluise in order to satisfy the enlarged size bound. Table \ref{tab:case_study} shows the detailed runtime information of SEA, where the community is refined iteratively (i.e., decreased $\delta^\star$ and $\varepsilon$) and finally the relative error is bounded by the default $e$=$2\%$. Besides, since we apply an error-based incremental sampling, we require a smaller $\Delta S$ than the initial $S$ to update the result.

\subsection{Parameter Sensitivity}
\label{sec:parameter}

\noindent\textbf{Effect of $\lambda$}. Figure \ref{fig:parameter}(a)-(b) show that the runtime increases as $\lambda$ increases, e.g., from 180 ms to 290 ms for DBLP. The more the samples, the more the time is required to greedy search communities from a larger $\tilde{H}_k$ for estimation. Besides, $\lambda$ has little effect on the attribute cohesiveness as the effectiveness is mainly dominated by the estimation with accuracy guarantee.

\vspace{0.1cm}
\noindent\textbf{Effect of $\epsilon$ and $1-\beta$ for Hoeffding Inequality}. Figure \ref{fig:parameter}(c)-(f) show that the response time increases as $\epsilon$ ($1-\beta$) decreases (increases). The stricter the $\epsilon$ and $1-\beta$, the more the nodes are required to form $G_q$ (Theorem \ref{th:hoeffding3}), leading to more time for estimation over a larger $\tilde{H}_k\subseteq G_q$. Since $\epsilon$ and $1-\beta$ are used to control the probability of the event that $G_q$ contains all nodes from the ground-truth community, the stricter the $\epsilon$ and $1-\beta$, the more the possibility of finding a better community.

\vspace{0.1cm}
\noindent\textbf{Effect of $e$ and $1-\alpha$ for BLB estimation}. Figure \ref{fig:parameter}(g)-(j) show that the stricter the $e$ and $1-\alpha$, the more the response time is required to achieve a smaller relative error. The relative error are almost bounded by $e$ except the case for $e$ = $2\%$ in Twitter. This is because Theorem \ref{th:relative_error} holds with a probability of $1-\alpha$ and this situation rarely happens for a large $1-\alpha$.

\vspace{0.1cm}
\noindent\textbf{Effect of $k$}. Figure \ref{fig:parameter}(k)-(l) show that $\delta(\cdot)$ increases as $k$ increases. A large $k$ usually indicates a small community, of which many nodes may important to the structure cohesiveness and a $k$-core would collapse if we delete such a node. So, the returned community may include some dissimilar nodes but contribute a lot to the structure cohesiveness, leading a larger $\delta(\cdot)$. The runtime for a small $k$ is usually more than that for a large $k$, as a small $k$ often indicates that we need more time for greedy search over a large $\tilde{H}_k$ for estimation.

\begin{figure}[t]
  \vspace{-0.15cm}
  \setlength{\abovecaptionskip}{0cm}
  \setlength{\belowcaptionskip}{0cm}
  \subfigure[DBLP]{
  \begin{minipage}{0.45\linewidth}
    \includegraphics[scale=0.38]{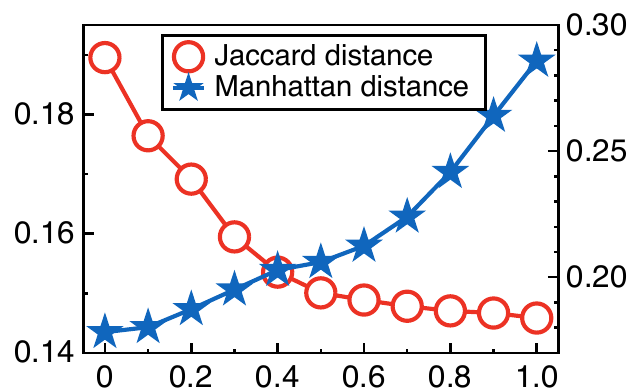}
    \vspace{-0.3cm}
  \end{minipage}
  }
  \hspace{0.25cm}
  \subfigure[Twitter]{
  \begin{minipage}{0.45\linewidth}
    \includegraphics[scale=0.38]{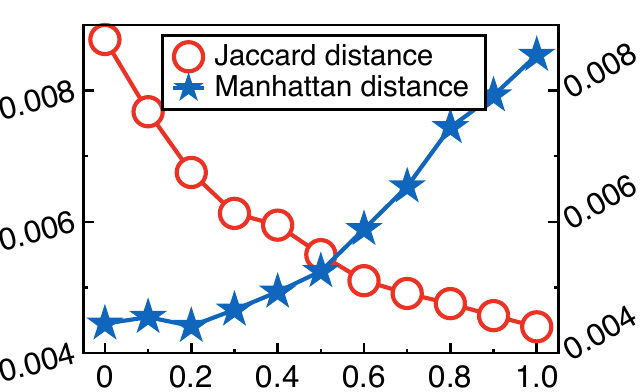}
    \vspace{-0.3cm}
  \end{minipage}
  }
  \caption{Effect of $\gamma$ on independent attribute cohesiveness}
  \label{fig:gamma}
  \vspace{-0.1cm}
\end{figure}

\vspace{0.1cm}
\noindent\textbf{Effect of $\gamma$}. Since $\gamma$ is a balance factor to adjust user preferences for two types of attribute cohesiveness, we varied $\gamma$ to study its effect on the two independent textual and numerical attribute cohesiveness in Figure \ref{fig:gamma}. We observed that when $\gamma=1$ ($\gamma=0$), our method tends to identify communities with the highest (lowest) cohesion in textual attributes (i.e., Jaccard distance) but the lowest (highest) cohesion in numerical attributes (i.e., Manhattan distance). A balance is achieved if $\gamma$ is close to 0.5, indicating that communities with good cohesion in both types of attributes can be identified.

\section{Related Work}
Community search (CS) was first studied in \cite{Sozio2010}, which can be divided into two categories according to graph types.

\noindent\textbf{CS on homogeneous graphs.} Many works focus on modeling the cohesive community based on minimum degree \cite{Sozio2010,Cui2014,Fang2019}, $k$-core \cite{batagelj2003m,Bonchi2019,Fang2019,Barbieri2015}, $k$-truss \cite{Akbas2017,Huang2014,Huang2015,Liu2020a}, $k$-clique \cite{Cui2013,Tsourakakis2013,Yuan2018}, $k$-edge \cite{Chang2015,Hu2017}, and query-biased density model \cite{Wu2015}. These works greatly boost the study of CS, but ignore the CS on attributed graphs. Thus, many works define different metrics of attribute cohesiveness, and then integrated it with the structure cohesiveness for CS \cite{Chen2018,Chen2019,Fang2017,Fang2017a,Fang2016,Huang2017,Liu2020,Zhang2019}. Although many metrics have been proposed, they are not strict enough to reflect a community's attribute cohesiveness. For example, \cite{Huang2017} measure a community's attribute cohesiveness as the weighted sum of each attribute's coverage, where coverage is computed as the ratio of nodes with exactly matched attribute. Similarly, \cite{Fang2016} uses \# shared attributes to measure cohesiveness, relying on equality matching too. Due to the constraints of equality matching, they are not well-suited for numerical attributes, for instance, it's more reasonable to seek similar movies with similar ratings rather than identical ones. \cite{Liu2020} aims to minimize the maximum attribute distance (i.e., optimize the worst case) in a community, but overlooks the similarity of nodes to the query node $q$. This motivates us to present a CS based on a $q$-centric attribute distance considering both textual and numerical attribute.


\vspace{0.1cm}
\noindent\textbf{CS on heterogeneous graphs.} Recently, CS on heterogeneous graphs has emerged. The meta-path $\mathcal{P}$ is often used to indicate the relation between two node types. Some community models are proposed, e.g., $(k,\mathcal{P})$-core \cite{Fang2020}, $(k,\mathcal{P})$-Btruss and $(k,\mathcal{P})$-Ctruss \cite{Yang2020}. Many follow-up works use them for various downstream applications, i.e., expert finding in \cite{Xu2022,Wang2022a} and influential community search via skyline influence vectors in \cite{Zhou2023}. \cite{Qiao2021} presents a keyword-centric CS, which takes a set of keywords as input rather than a query node, and it cannot support numerical attributes. It ensures that any node in a community can reach to a keyword with a shorter path, rather than all nodes in a community are similar in their attributes. 

None of above methods provide an efficient approximate solution with a reliable evaluation on community's quality based on a metric that can better distinguish a community’s attribute cohesiveness, inspiring our study in this paper.


\vspace{-0.1cm}
\section{Conclusions}
We study an NP-hard CS-AG problem atop a strict $q$-centric attribute cohesiveness metric for $k$-core model on homogeneous graphs. We first propose an exact method with three pruning strategies served as a baseline. Then, we propose a sampling-estimation-based method to quickly return an appropriate community with an accuracy guarantee (given as an error-bounded confidence interval). We extend our method to heterogeneous graphs, $k$-truss model, and size-bounded CS. Experimental studies on ten real-world datasets demonstrate our method's superiority in both effectiveness and efficiency.

\bibliographystyle{IEEEtran}
\bibliography{CS}

\end{document}